# A qualitative and quantitative analysis of open citations to retracted articles: the Wakefield *et al.*'s case

Ivan Heibi – ivan.heibi2@unibo.it – https://orcid.org/0000-0001-5366-5194
Research Centre for Open Scholarly Metadata, Department of Classical Philology and Italian Studies, University of Bologna, Bologna, Italy
Digital Humanities Advanced Research Centre (/DH.arc), Department of Classical Philology and Italian Studies, University of Bologna, Bologna, Italy

Silvio Peroni – silvio.peroni@unibo.it – https://orcid.org/0000-0003-0530-4305
Research Centre for Open Scholarly Metadata, Department of Classical Philology and Italian Studies, University of Bologna, Bologna, Italy
Digital Humanities Advanced Research Centre (/DH.arc), Department of Classical Philology and Italian Studies, University of Bologna, Bologna, Italy

## Abstract

In this article, we show the results of a quantitative and qualitative analysis of open citations on a popular and highly cited retracted paper: "Ileal-lymphoid-nodular hyperplasia, non-specific colitis, and pervasive developmental disorder in children" by Wakefield *et al.*, published in 1998. The main purpose of our study is to understand the behavior of the publications citing retracted articles and the characteristics of the citations the retracted articles accumulated over time. Our analysis is based on a methodology which illustrates how we gathered the data, extracted the topics of the citing articles, and visualized the results. The data and services used are all open and free to foster the reproducibility of the analysis. The outcomes concerned the analysis of the entities citing Wakefield *et al.*'s article and their related in-text citations. We observed a constant increasing number of citations in the last 20 years, accompanied with a constant increment in the percentage of those acknowledging its retraction. Citing articles have started either discussing or dealing with the retraction of Wakefield *et al.*'s article even before its full retraction, happened in 2010. Articles in the social sciences domain citing the Wakefield *et al.*'s one were among those that have mostly discussed its retraction. In addition, when observing the in-text citations, we noticed that a large part of the citations received by Wakefield *et al.*'s article has focused on general discussions without recalling strictly medical details, especially after the full retraction. Medical studies did not hesitate in acknowledging the retraction and often provided strong negative statements on it.

## Introduction

A peer-reviewed retracted article should be considered as an invalid source of knowledge depending on specific reasons for its retraction which might include scientific misconduct, fabrication, general content errors, plagiarism and self-plagiarism (Moylan & Kowalczuk, 2016). The editor(s) of the venue in which the original publication was published has the final decision about the retracting material. This decision is accompanied by a retraction notice. Also, sometimes a label (e.g. "RETRACTED") is associated with the retracted article either in the article title or as a label specified upon the article content.

In order to make retractions more visible to a reader, existing services, such as CrossMark by Crossref, have been proposed and implemented in the past years to show updated notifications regarding articles, such as retractions and error corrections (Meyer, 2011). An important service which keeps track of and collects retractions of scholarly articles is Retraction Watch (http://retractionwatch.com/) (Collier, 2011). In addition, the COPE retraction guidelines (Barbour *et al.*, 2009) states that "the original article should not be completely removed or 'replaced', but should be retained and linked to", practically enabling studies on the retracted articles.

The retraction phenomenon has been largely discussed by scientometricians. We can organize the studies in this domain into two macro categories: (a) large scale analysis and (b) case study analysis.

Works belonging to category (a) focus on either an analysis of a single field of study or a broader domain, such as a macro area. Usually, these studies try to answer general questions such as how retractions influence the impact on the authors, institutions and the retracted work itself.

Large scale citation analysis on retracted articles have been mostly focused on quantitative aspects. For instance, by considering the reasons for retraction introduced in (Bar-Ilan *et al.*, 2018) and (Lu *et al.*, 2013), authors used the citation data collected from Web of Science to demonstrate that a single retraction could trigger citation losses through an author's prior body of work. The negative repercussions on authors and co-authors of retracted articles have been shown also by other works such as (Azoulay *et al.*, 2017) (Mongeon & Larivière, 2016) (Shuai *et al.*, 2017). Along the same lines, Feng *et al.* (2020) introduced a multi-dimensional observation framework using four dimensions, including scientific impact, technological impact, funding impact and Altmetric impact. Others have discussed possible approaches to avoid retraction and related issues in citing retracted papers. For example, Mott *et al.* (2019) have suggested strategies to adopt for improving the effectiveness of retraction notices and the awareness of the citing authors. Another example is the work done by Bordignon (2020) which investigated the different impacts that negative citations in articles and comments posted on post-publication peer review platforms have on to the correction of science. Finally, a recent work has been done by Bolland *et al.* (2021), in which the authors focused on the analysis of the citations made before the retraction.

The studies of category (b) consider either single or multiple retracted article cases (usually, popular cases) and perform content analysis of the articles citing retracted ones. Generally, their main goal is building a general approach to apply on a larger scale corpus starting from the findings and results obtained – e.g. by focusing on post-retraction citations and the related sentiment when citing (Bar-Ilan & Halevi, 2017), by classifying citation contexts (Jan *et al.*, 2018), and by running a network analysis study (Chen & Leydesdorff, 2014). Similarly, the work done by Teixeira da Silva & Dobránszki (2017) focused on a restricted list of the top ten cited retracted articles for analyzing the number of citations and retraction reasons, without considering the content of the articles citing the retracted ones.

Other studies of category (b), instead, focus only on one specific retraction case. The aims of these works are to observe where in-text citations to the retracted article appear in the text of the citing articles in order to perform a network analysis which monitors the propagation of the results and findings of the retracted article (van der Vet & Nijveen, 2016), to notice whether retracted works are still being cited without mentioning their retraction (Bornemann-Cimenti *et al.*, 2016), and to classify reasons for citing the retracted articles (Luwel *et al.*, 2019). The work done by Schneider *et al.* (2020) is another example of a work falling into this category of studies.

The analysis we present in this article is close to those introduced in the latter set of studies. We want to focus on a highly cited retracted article, i.e. (Wakefield *et al.*, 1998), that suggested a link between autism and childhood vaccines. This article was partially retracted in 2004, and subsequently fully retracted in 2010. Throughout our article, we refer to it with the abbreviation WF-PUB-1998.

We think that the retraction of WF-PUB-1998 is an important case that deserves to be analyzed considering its popularity among several anti-vaccine movements and the implications it has had for society (Chen *et al.* 2013). Since WF-PUB-1998 is also one of the most cited retraction cases, this large quantity of citations will help us have a better assessment of the methodology we introduce in this work.

In our study, we focused on the citation analysis of the WF-PUB-1998 from a quantitative and qualitative point of view. We split the analysis and the findings into three different periods – i.e., P1, P2, and P3, based on the years of the partial (2004) and final (2010) retraction of WF-PUB-1998. In particular P1 refers to the period from WF-PUB-1998 publication to the partial retraction (1998-2004), P2 from the year after the partial retraction to the full retraction (2005-2010), and P3 from the year after the final retraction to 2017 (2011-2017). We considered 2017 as last year due to the availability of the citation data in OpenCitations (Peroni & Shotton, 2020) we gathered at the time this work was performed, i.e., the November 2018 COCI release (OpenCitations, 2018). Our workflow follows a precise methodology specifically designed for the application of a qualitative and quantitative analysis on retracted articles. The methodology goes through three main stages: annotation, data and results visualization, and answering the research questions.

A similar work on the retraction case of WF-PUB-1998 has been recently also presented by Suelzer *et al.* (2019). The analysis of Suelzer *et al.* is based on a collection of 1,153 citing works retrieved using Web of Science and the findings are also compliant with the three periods P1-P3 we have used in our work. The collected citations have been classified into negative/perfunctory/affirmative and annotated as those that have/have not documented the partial/full retraction of WF-PUB-1998. Suelzer *et al.* suggest that improvements are needed from publishers, bibliographic databases, and citation management software to ensure that retracted articles are accurately documented.

Unlike (Suelzer *et al.*, 2019), our work relied on open and free services to retrieve the articles (and their metadata) citing WF-PUB-1998, and we used automatic natural language processing techniques to conduct a qualitative study on the content of the article citing WF-PUB-1998. We present a detailed comparison that highlights the method, annotated features, and findings of our work and (Suelzer *et al.*, 2019) in the final part of this article.

The aim of our work is to answer to the following research questions:

- RQ1) What are the research topics introduced in the articles citing WF-PUB-1998 before and after its retraction?

- RQ2) What are the most relevant characteristics of the in-text citations (e.g. intent, sentiment, mention of the retraction, etc.) in the articles citing WF-PUB-1998 before and after its retraction?

# Methodology

The methodology of this work is based on three different steps. The first two steps (subsections "Data gathering" and "Topic modeling") define the methods for annotating and generating the data that we need for our study, while the third step (subsection "Addressing the research questions") defines how we try to answer RQ1 and RQ2.

Before describing the steps of our methodology, though, we give a preliminary and brief introduction to open citations, since they represent our main source to gather initial citation data. In particular, we provide their definition and usage, since they represent an important part of our methodology.

## Open citations

Following the definition provided in (Peroni & Shotton, 2018a), a bibliographic citation is an open citation when the data needed to define the citation are:

- Structured: expressed in one or more machine-readable format such as JSON.
- Separate: available without the need to access the source article in which the citation is defined.
- Open: freely accessible and reusable without restrictions.
- Identifiable: the entities linked by an open citation must be clearly identified by using a specific persistent identifier scheme, such as a DOI, or a URL.
- Available: it must be possible to obtain the basic metadata of the entities involved in the citation by resolving their identifiers.

The open citation data used in this work are provided by OpenCitations (Peroni & Shotton, 2020), an independent infrastructure organization for open scholarship, which is dedicated to the publication of open citation data using Semantic Web technologies (Berners-Lee *et al.*, 2001), which facilitate the use of precise semantics for the encoding and creation of machine-processable data on the Web.

The Semantic Web technologies used by OpenCitations permit the publication of bibliographic and citation data as Linked Open Data (LOD) (Bizer *et al.*, 2009). These bibliographic and citation data are compliant with the OpenCitations Data Model (Daquino *et al.*, 2020), which is implemented by means of the SPAR Ontologies (http://www.sparontologies.net) (Peroni & Shotton, 2018b). In particular, citations are described using Citation Typing Ontology (CiTO, http://purl.org/spar/cito) (Peroni & Shotton, 2012), which allows one to create metadata describing citations (that are distinct from the metadata describing the cited works themselves) and permits the *intent* of an author when referring to another document to be captured.

In particular, the OpenCitations collection we used to gather all the open citation data is COCI, the OpenCitations Index of Crossref open DOI-to-DOI citations (http://opencitations.net/index/coci) (Heibi *et al.*, 2019b), which contains details of all the citations that are specified by the open references to DOI-identified works present in Crossref (Hendricks *et al.*, 2020).

COCI and CiTO are two of the main components in the methodology we present. We use COCI to gather the citations of the retracted article in consideration, and we adopt the CiTO definitions to characterize the citation intents, based on the citation context and following a guiding schema.

## Data gathering

The data sources we used to gather the data for our analysis were OpenCitations COCI, that we used to retrieve citation data, the RetractionWatch database (http://retractiondatabase.org/) used to retrieve information of retracted articles, SCImago (https://www.scimagojr.com/) to retrieve subject areas and subject categories of articles, and the ISBNDB service (https://isbndb.com/) to look up the Library of Congress Classification code (LCC, https://www.loc.gov/catdir/cpso/lcco/) of books.

We queried the COCI REST API (http://opencitations.net/index/coci/api/v1) when COCI was populated with citation data from its November 2018 release (OpenCitations, 2018), that contained 445,826,118 citation links coming from 46,534,705 bibliographic resources. Among the attributes that COCI uses for characterizing each citation having WF-PUB-2018 as cited entity, we took into consideration the citing DOI, the cited DOI, and the creation date of the citation (i.e. the publication date of the citing entity).

This stage was organized in five steps, introduced in Table 1. Additional information and details about the step related to the gathering of the data used in our analysis can be found in (Heibi & Peroni, 2020), which goes deeper into its technical aspects (e.g. execution of software codes and additional contextual information) and does not discuss any aspect related with the other steps of the methodology. Thus, the methodology described herein is self-contained and enables the reproducibility of our work.

### Gathering raw data

We retrieved the DOI, year of publication, title, ISSN/ISBN of the publication venue and the related title of all the articles citing WF-PUB-1998 starting from its DOI. For doing that, we used the "citations" operation of the OpenCitations COCI API (http://opencitations.net/index/coci/api/v1#/citations/{doi}) to get the list of all citing entities, then we used the "metadata" operation (http://opencitations.net/index/coci/api/v1#/metadata/{dois}) to get the metadata of each citing entity.

**Table 1.** An overview of all the steps needed for generating an annotated dataset of WF-PUB-1998 citing entities, to be further used during this work. For each step, we provide a brief description, the inputs needed, and the output produced. The output is represented as the list of features that will be included in the final dataset used for our analysis.

| Step | Description | Input | Output |
|---|---|---|---|
| 1) Identifying and retrieving the citing entities | Identifying the list of entities citing WF-PUB-1998 and storing their main metadata | DOI of the retracted article | For each citing entity:<br>*1.1) DOI*<br>*1.2) year of publication*<br>*1.3) title*<br>*1.4) venue id (ISSN/ISBN)*<br>*1.5) venue title* |
| 2) Retrieving the citing entities characteristics | Annotating whether the citing entities have been or have not been retracted as well | DOIs of the citing entities | For each citing entity:<br>*2.1) is / is not retracted* |
| 3) Classifying the citing entities according to subject areas and subject categories | Classifying the citing entities into areas of study and specific subject categories, following the SCImago classification | ISSN/ISBN of publication venues of citing entities | For each citing entity:<br>*3.1) subject area*<br>*3.2) subject category* |
| 4) Extracting textual values from the citing entities | Extracting the citing entities' abstracts, the in-text reference pointers, citation contexts, title of the section where the in-text citations happen | DOIs of the citing entities | For each citing entity:<br>*4.1) abstract*<br>*4.2) in-text citation section*<br>*4.3) in-text citation context*<br>*4.4) in-text reference pointer* |
| 5) Annotating the in-text citations characteristics | Manually annotating the intent (based on the citation functions of CiTO) and sentiment of each in-text citation, and specifying whether the text in citation contexts mentions the retraction of the cited article | In-text citation contexts | For each in-text citation:<br>*5.1) citation intent*<br>*5.2) citation sentiment*<br>*5.3) retraction is / is not mentioned* |

Then, we queried the RetractionWatch database to manually check if each of the citing entities (identified by its DOI) has been retracted as well or not, and we identified the subject areas and subject categories of each citing entity using the identifiers (either ISSN or ISBN) of the publication venue of the cited entity. For publication venues with ISSN, we used the SCImago Journal Rank. SCImago groups the journals into subject areas (27 major thematic areas, e.g. *Medicine*) and subject categories (313 specific subject categories, e.g. *Immunology and Allergy*). Some venues can have more than one subject area and subject category – we considered all of them in these cases. For publication venues with ISBN (mainly books), we used the ISBNDB service to look up the related Library of Congress Classification code. Then, we mapped the LCC categories we found to SCImago subject areas and categories as follows:

1. We considered only the starting alphabetic segment of the LCC code and find the corresponding LCC discipline using a pre-built lookup index (e.g. "*RC360*" -> "*RC*" -> "*Medicine*").

2. We checked whether the value of the LCC subject matches the exact value of a Scimago area using a pre-built Scimago mapping index, which is available at https://github.com/ivanhb/cits-ret-method. If this is true, we automatically annotated the subject area with such value, while we assigned, as subject category, the same value with the addition of "(miscellaneous)" at the end of it. This is usually done on the Scimago classification to express a general category of a specific area of study. In case no corresponding Scimago area has been found, we continued to point 3.
3. We checked whether the value of the LCC subject is a Scimago category using the same pre-built Scimago mapping index. If this is the case, we annotated the corresponding category with such value, while the area will have the same value used on the Scimago classification to denote the macro area of such category. In case no corresponding Scimago category was found, we continued to point 4.
4. The remaining ISBN values needed to be manually annotated through the consultation of the complete LCC index (http://www.loc.gov/catdir/cpso/lcco/).

Finally, starting from the DOI of the citing entities, we retrieved the full-text of all the citing articles. From the full-text of such articles, we extracted their abstracts, the in-text reference pointers denoting a bibliographic reference referencing WF-PUB-1998 (e.g. "Wakefield *et al.*, 1998"), the citation contexts of the in-text citations and the sections where the citation contexts were contained.

For this study, we defined the in-text citation contexts as the sentence that contains the in-text reference pointer to WF-PUB-1998 (i.e. the anchor sentence), plus the preceding and following sentences. As suggested in (Ritchie, Robertson, and Teufel 2008), this strategy for the definition of the citation context window seemed appropriate to guarantee an accurate annotation of the intents of citations. Special cases/exceptions to this rule (e.g. if the sentence that contains the in-text reference is the first sentence of a section, then the in-text citation context did not include the preceding sentence) are treated in (Heibi & Peroni, 2020). Also, we characterized each of the sections containing in-text citations according to their type – using the categories "introduction", "method", "abstract", "results", "conclusions", "background", and "discussion" listed in (Suppe, 1998). These categories have been used when the intent of the section was clear, otherwise we used other three residual categories, i.e. "*first section*", "*final section*" and "*middle section*" combined with the original title of the section. If the examined full-text of the citing entity is not organized into sections/paragraphs, then the value of its in-text citation section is set to "*none*". For instance, this could be the case for citing entities that are editorials.

## Annotating the in-text citations

We analyzed each citation context of the in-text citation retrieved, and we inferred:

- the perceived sentiment regarding WF-PUB-1998;
- whether at least one citation context of any in-text citation of the citing entity does explicitly mention the fact that the cited entity has been retracted (i.e. the citation context contained the word "retract" or one of its derivative words – "retractions", "retracted", etc.);
- the citation intent (or citation function), defined as the authors' reason for citing a specific article (e.g. the citing entity uses a method defined in the cited entity).

For specifying the citation sentiment, we followed the classification proposed by (Bar-Ilan & Halevi, 2017). Thus, we annotated each in-text citation with one of the following values:

- positive, when the retracted article was cited as sharing valid conclusions, and its findings could have been also used in the citing study;
- negative if the citing study cited the retracted article and addressed its findings as inappropriate and/or invalid;
- neutral, when the author of the citing article referred to the retracted article without including any judgment or personal opinion regarding its validity.

To record the citation intent, we used the citation functions specified in the Citation Typing Ontology (Peroni & Shotton, 2012). Even if, in principle, an in-text citation might refer to more than one CiTO function at the same time, we decided to annotate each in-text citation with one citation function only. In our methodology, we made a clear distinction between the *sentiment* and the *intent* of the citation, since the annotation of a specific citation intent does not directly imply its sentiment. For instance, the intent might be *to obtain background from* the cited entity, yet this could be done with a negative/positive perception toward it. For instance, the authors of the following in-text citation cited WF-PUB-1998 to obtain background information from it, and they expressed a slightly negative sentiment toward it: "We explain one example of single-source overlays in detail here. The seed article in the example, Wakefield *et al.* (1998), is a highly cited retracted article, which has profound implications on public health, especially on vaccine uptakes from children" (Gap Analytics, 2014).

We performed a manual annotation of the in-text citations using the decision model we developed for this study, summarized in Figure 1. The decision model is organized into three main macro categories (i.e., large columns): (1) "Reviewing …", (2) "Affecting …", and (3) "Referring …". Each macro category has at least two other inner classifications (i.e., columns). For instance, the "Affecting" macro-category has the "citing entity" and "cited entity" inner columns. The macro categories and their inner classifications work as guiding schemas for the annotator and are not part of the final annotations.

The decision model is based on a priority ranked strategy that works as follows:

1. we matched each in-text citation to WF-PUB-1998 against at least one of the three macro-categories, i.e. "Reviewing", "Affecting" and "Referring" (first row in Figure 1);
2. for each macro-category selected, we selected one or more citation functions choosing between those provided by CiTO – the decision model provides a template and an example (i.e., *"Fill the sentence …"*) to help us chose the most suitable one;
3. if we selected only one citation function, we annotated the in-text citation intent with that function; otherwise
4. we calculated the priority of each citation function selected by summing its value in parenthesis (from 0.1 to 0.6) with the corresponding value defined in the x-axis (from 1 to 8) and in the y-axis (from 10 to 50), as shown in Figure 1. The smaller the sum, the more priority the citation function has. For instance, the priority of the citation function "confirms" is 11.2 that is higher than the one of the citation function "describes", which is 43.2. Finally, we selected the citation function that has higher priority and annotated the in-text citation function with it.

|  | Reviewing and eventually giving an opinion on the cited entity<br><br>*Fill in the sentence:*<br>"My statements are  _HEADER_  the cited entity, such that they  _CiTO-citation-function_ "<br><br>*E.g.* "My statements are  _Not on the same page with_  the cited entity, such that they  _critiques_ " | | | Affecting either the content of or the perception toward the cited/citing entity<br><br>*Fill in the sentence:*<br>"My statements  _CiTO-citation-function_  the cited entity , and affect the content of/perception toward the  _HEADER_ "<br><br>*E.g.* "My statements  _corrects_  the cited entity , and affect the content of/perception toward the  _Cited entity_ " | | Referring to the cited entity for material/conceptual purposes<br><br>*Fill in the sentence:*<br>"The document I am citing represents a  _HEADER_ , such that my statements  _CiTO-citation-function_  the cited entity"<br><br>*E.g.* "The document I am citing represents a  _General source_ , such that my statements  _cites for information_  the cited entity" | | |
|---|---|---|---|---|---|---|---|---|
|  | Consistent with | Inconsistent with | Talking about | Cited entity | Citing entity | Material | Concept | General source |
| 10 | (0.1) supports<br>(0.2) confirms | (0.1) derides<br>(0.2) ridicules<br>(0.3) refutes<br>(0.4) critiques | | | | | | |
| 20 | (0.1) agrees with | (0.1) disagrees with<br>(0.2) disputes | | (0.1) compiles<br>(0.2) retracts<br>(0.3) replies to<br>(0.4) speculates on<br>(0.5) corrects<br>(0.6) extends | (0.1) uses data from<br>(0.2) uses method in<br>(0.3) uses conclusions from<br>(0.4) obtains support from | | | |
| 30 | | | (0.1) parodies<br>(0.2) qualifies<br>(0.3) credits | (0.1) updates | (0.1) obtains background from | | | |
| 40 | | | (0.1) discusses<br>(0.2) describes | | (0.1) includes quotation from | | | |
| 50 | | | | | (0.1) includes excerpt from<br>(0.2) documents<br>(0.3) reviews | (0.1) cites as metadata document<br>(0.2) cites as data source<br>(0.3) cites as source document | (0.1) cites as authority<br>(0.2) cites as evidence<br>(0.3) cites as potential solution<br>(0.4) cites as recommended reading<br>(0.5) cites as related | (0.1) cites for information |
| Score | 1 | 2 | 3 | 4 | 5 | 6 | 7 | 8 |

**Figure 1**. The decision model for the selection of a CiTO citation function to use for the annotation of the citation intent of a an examined in-text citation based on its context. The first large row contains the three macro-categories: (1) "Reviewing …", (2) "Affecting …", and (3) "Referring …". Each macro-category has at least two subcategories, and each subcategory refers to a set of citation functions. The first row defines what are the citation functions suitable for it through the help of a guiding sentence which needs to be completed according to the chosen sub-category and citation function.

# Topic modeling

Some recent works such as (Bornmann *et al.*, 2020) and (Crothers *et al.*, 2020) analyzed the context of citations to highly cited articles that have explained and introduced important concepts. The idea of these works is to count the number of times the concepts are mentioned in the citation context. This analysis only makes sense if the highly cited publications have introduced at least one important concept. The work done by Lyu and Costas (2020) examined the Big Data research domain and investigated how the academic topics shift across altmetric sources (e.g. Twitter). Another recent work proposed by Zhang *et al.* (2021) analyzed the topic evolution in early COVID-19 research. Generally, an analysis toward the topics evolution in specific research domains, institutions, periods, or following an important event (e.g., the COVID-19 pandemic), is an important subject of interest for the development of effective ways to inform research strategies and evaluate research activities, as it is demonstrated by the development of tools such as Elsevier's SciVal (https://www.elsevier.com/solutions/scival), a web-based tool for visualizing and investigating these aspects.

In our study, we wanted to generalize to go beyond a set of popular concepts and we tried to consider to an arbitrary number of concepts/topics that we want to identify using a computational approach. We decided to address this problem by using a topic modeling technique, which is an appropriate method to use for the automatic analysis of texts that works without having preliminary knowledge on the subjects the texts are about. For instance, a recent application of topic modelling for similar purposes is described in the work by Han (2020), who used it to investigate the evolution of research topics in the library and information science (LIS) domain.

We run a topic modeling analysis on the textual features we gathered (i.e. abstracts and citation contexts) using MITAO (https://github.com/catarsi/mitao) (Ferri *et al.*, 2020), a visual interface to create a customizable visual workflow based on the Latent Dirichlet Allocation (LDA) topic modeling (Jelodar *et al.*, 2019). In particular, the topic modeling analysis we introduce in this article resulted in the creation of two topic models, one for the abstracts and one for the in-text citation contexts. This analysis is based on three main stages: (a) the identification of the number of topics to consider given a corpus of texts, (b) building the topic modeling workflow, and (c) generating the results and the related visualizations. We discuss each of these stages individually in the following subsections.

## Number of topics

To decide about the right number of topics to consider in each case, we computed and used the topic coherence score, as suggested in (Schmiedel *et al.*, 2019). This score measures the degree of semantic similarity between high scoring words in the topic, and it helps us distinguish between topics that are semantically interpretable and topics that are artifacts of a mere statistical inference. Thus, for each of our cases (abstracts and citation contexts), we calculated the average coherence score for a range of models trained with a different number of topics (from 1 topic to 40 topics). Then, we plotted these values, we observed the number of topics for which the average score plateaued, and we selected a number of topics indicated in the plateau.

Figure 2 shows the coherence score values of different LDA topic models built with a number of topics ranging from 1 to 40 using the citation contexts. The coherence score plateaued around 22-23 topics. Thus, we decided to consider 22 topics for the citation contexts. We have used a similar approach for abstracts.

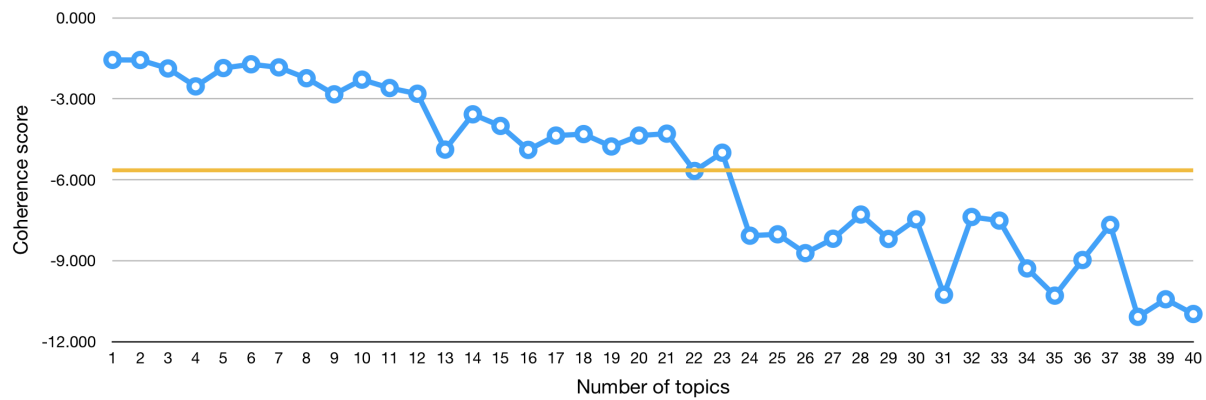

**Figure 2.** The coherence score of different LDA topic models built using a variable number of topics, from 1 to 40. The topic model is based on the corpus and dictionary of the in-text citation contexts. The orange line is the average value, and it plateaus around 22-23 topics.

## The topic modeling workflow

A standard workflow for building a topic model is composed of three main steps. Tokenization is the process of converting the text into a list of words, by removing punctuations, unnecessary characters, and stopwords. In our study, stopwords also included, for abstracts, tokens used in structured abstracts such as "background", "summary", and "results", and, for citation contexts, tokens used in the bibliographic reference of WF-PUB-1998 such as "Wakefield", "Ileal", and "lymphoid". Since topic modeling can drastically benefit from the lemmatization (May *et al.*, 2019), we decided to lemmatize all the tokens obtained by processing the abstracts and citation contexts.

Then, we created vectors for each of the tokens retrieved. In particular, we used the term frequency-inverse document frequency (TF-IDF) model to vectorize our words. The TF-IDF model takes into account the importance of the words based on its rarity in the document (i.e. either the abstract or the citation contexts). This model is considered as a good word weighting schema for general purpose textual collections and when the frequent terms may not be that representative of the document topics (Bengfort *et al.*, 2018) (Truica *et al.*, 2016).

Finally, we built two topic models, one for the abstracts and one for the in-text citation contexts and we gave as input the number of topics, identified using the coherence score, to each model following the results of the previous stage. Figure 3 illustrates graphically the workflow we developed and run using MITAO. Some of the components of the workflow are used to generate the results and the visualizations, which are introduced in the following subsection.

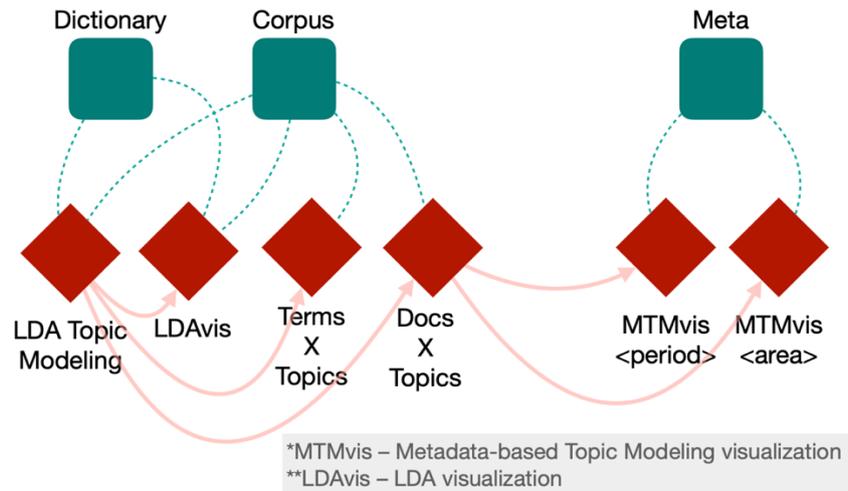

**Figure 3.** The workflow, created via MITAO, we used for computing the LDA topic modeling and generating the LDAvis (LDA visualization) and MTMvis (Metadata-based Topic Modeling visualization) visualizations (the tools "LDAvis", "MTMvis <period>", and "MTMvis <area>"). The green squares are used to specify input material which is considered by the various tools composing the workflow (i.e., the red rhombi). In particular, the workflow takes three inputs: (a) the vectorized corpus ("Corpus"), (b) a dictionary of words based on the tokenization results ("Dictionary"), and (c) the metadata of the original documents forming the corpus ("Meta"). The arrows between the tools indicate the direction of the data flow and the output-input relation among them. For instance, the execution of the workflow starts with the tool "LDA Topic Modeling", that takes in input the "Corpus" and the "Dictionary" and produces an output that is used as part of the input for other three tools, i.e. "LDAvis", "Terms X Topics" and "Docs X Topics".

## Results and visualizations

As anticipated in the previous subsections, we used MITAO to generate two datasets for each case (abstracts and citation contexts). Each dataset contained:

- the 30 most important keywords of each topic, which represent the 30 most useful and probable terms for interpreting a topic, ranked according to their probability value;
- documents representativeness, i.e. the lists of all the documents of the corpus and their representativeness against each topic.

We also used MITAO for generating two interactive visualizations which we used to highlight important aspects of our study: LDAvis and MTMvis.

LDAvis provides a graphical overview of the topic modeling results (Sievert & Shirley, 2014). This visualization plots the topics as circles in a two-dimensional plane whose centers are determined by computing the distance between topics and uses a multidimensional scaling to project the inter-topic distances onto two dimensions. The topic prevalence is represented by the dimension of the area of each circle. LDAvis shows a global list of 30 terms ranked using the "term saliency" measure. This saliency measure combines the overall probability of a term with its distinctiveness: how informative is a specific term for determining the generation of a topic versus any other randomly selected term (Chuang *et al.*, 2012). In addition, one can select a singular topic and LDAvis will show a list of 30 terms ranked using the "relevancy" measures. We used the default relevancy metric as defined in

(Sievert & Shirley, 2014) to show the ranking of terms according to their topic-specific probability.

MTMvis (Metadata-based Topic Modeling Visualization) provides an interactive visualization which shows the representativeness of the topics in the documents based on a customizable metadata set specified for those documents. We created two visualizations for both the abstracts and the citation contexts based, respectively, on the year of publication and the subject area of the articles citing WF-PUB-1998.

## Addressing the research questions

To answer RQ1 and RQ2, we investigated the citing entities data and combined such data with the results of the topic modeling process.

Our approach takes into consideration the years of the partial (2004) and final (2010) retraction of WF-PUB-1998 to define three periods: (P1) from WF-PUB-1998 publication to the partial retraction (years 1998-2004), (P2) from the partial retraction and to the final retraction (years 2005-2010), and (P3) from the final retraction to 2017 (years 2011-2017).

We used LDAvis and MTMvis to analyze the results of the topic model obtained. On the one hand, we used LDAvis to have a general overview of the topics, inspect their prevalence, and their terms. On the other hand, we used MTMvis to plot the corpus documents' topic representativeness.

Regarding RQ1, we mainly needed to analyze the results obtained by the topic model of the abstracts. The idea was to monitor the evolution of the emerging topics considering the three periods P1-P3 to show the main arising changes. We compared these observations against the area of study to highlight the evolution of citing behavior in each individual area.

When dealing with RQ2, we primarily considered the features which characterized the in-text citations, such as the citation intent and the sentiment. The idea was to analyze these features against the outcomes of the topic model of the in-text citation contexts.

# Results

In this section, we present the results of our analysis. All the data and visualizations are available in (Heibi & Peroni, 2020). Although in this article we present a screenshot of the visualizations, these are provided in dynamic HTML documents, and each visualization can be customized using the filters and parameters it makes available. We provide a dedicated webpage (https://ivanhb.github.io/ret-analysis-wakefield-results/) to enable readers to use such dynamic visualizations that we present in this work.

We organize the presentation of the results in two sections describing (a) the entities citing WF-PUB-1998 and (b) their in-text citations to WF-PUB-1998. For both, we introduce the data and the features we used for the analysis and then we present the outcomes of the related topic models.

## Citing entities

The total number of citing entities gathered is 615. In Table 2, we list all the features we collected related to the citing entities. In particular, the first column lists the features with a brief description, while the second column summarizes its values, the total number of citing entities having such values and, if applicable, a classification of the different possible values.

Figure 4 introduces some descriptive statistics of the values described in Table 2. The charts are organized in three distinct rows, one for each period considered (P1-P3), mentioned in the first column. The second column contains the distribution per year of the citing articles according to the fact they either mention the retraction of WF-PUB-1998 (in green) or they do not (in red). On top of each bar in the chart, we also specify the number of citing entities the bar refers to. The third column contains the subject areas of the citing entities. The chart shows the ten most represented areas of study, while it groups all the other values (if any) in the last slice of the pie with the "Others" label.

**Table 2.** The features that directly characterize the citing entities. The first column lists the features with a brief description, while the second column summarizes the related values we gathered.

| WF-PUB-1998 citing entities features | Values |
| --- | --- |
| *doi* <br> The DOI of the citing article | **Total**: All the citing entities had a value specified. |
| *year* <br> The year of publication of the citing article | **Total**: All the citing entities had a value specified. <br> **Values:** From 1998 (year of publication of WF-PUB-1998) to 2017. |
| *title* <br> The title of the citing article | **Total**: All the citing entities had a value specified. |
| *source_id* <br> The ID (ISSN/ISBN) of the venue of publication of the citing article | **Total**: 599 (97%) citing entities had a value specified. <br> **Values:** ISSNs (548), ISBNs (51). |
| *source_title* <br> The title of the venue of publication of the citing article | **Total**: 603 (98%) citing entities had a value specified. |
| *retracted* <br> A yes/no value depending on whether the citing article has or has not received at least one retraction notification. | **Total**: 1 citing entity. |
| *area* <br> The subject areas of the venue of publication of the citing article, based on the the SCImago Journal Classification (https://www.scimagojr.com/) | **Total**: 576 (93%) citing entities had at least a value specified. <br> **Values:** 24 different values: "medicine" (380), "social sciences" (90), "nursing" (81), "biochemistry, genetics and molecular biology" (59), "psychology" (58), "pharmacology, toxicology and pharmaceutics" (54), "immunology and microbiology" (52), "arts and humanities" (28), "neuroscience" (24), "environmental science" (17), "agricultural and biological sciences" (16), "health professions" (15), "computer science" (13), "mathematics" (10), "business, management and accounting" (8), "engineering" (7), "dentistry" (7), "multidisciplinary" (7), "decision sciences" (7), "economics, econometrics and finance" (5), "earth and planetary sciences" (1), "chemical engineering" (1), "materials science" (1), "physics and astronomy" (1) |
| *category* <br> The subject categories of the venue of publication of the citing article, based on the the SCImago Journal Classification (https://www.scimagojr.com/) | **Total**: 576 (93%) citing entities had a value specified. <br> **Values**: 170 different values. |
| *abstract* <br> The abstract of the citing article | **Total**: All the citing entities had a value specified. |
| *mention_retraction* <br> A yes/no value that indicates if at least one of the citation contexts of the citing article explicitly mentions the fact that the cited entity is retracted. | **Total**: All the citing entities had a value specified. <br> **Values**: *no* (464), *yes* (151) |

The second column of Figure 4 shows a continuous increment in the number of citations and a higher percentage of entities mentioning the retraction. In 2009, we had the smallest percentage of entities which have mentioned the retraction (7%), while we observed the higher percentage value in 2017 (61%). Considering the distribution of the areas of study, we observed a slightly decreasing presence of the *medicine* area in favor of other areas of study which gained much more relevancy in P2 and P3 (e.g. *social sciences* being 1.61%, 8.93%, and 12.59% in P1, P2, and P3 respectively). In addition, we noticed the emerging of new areas in P2 and P3, such as *economics* and *environmental science*.

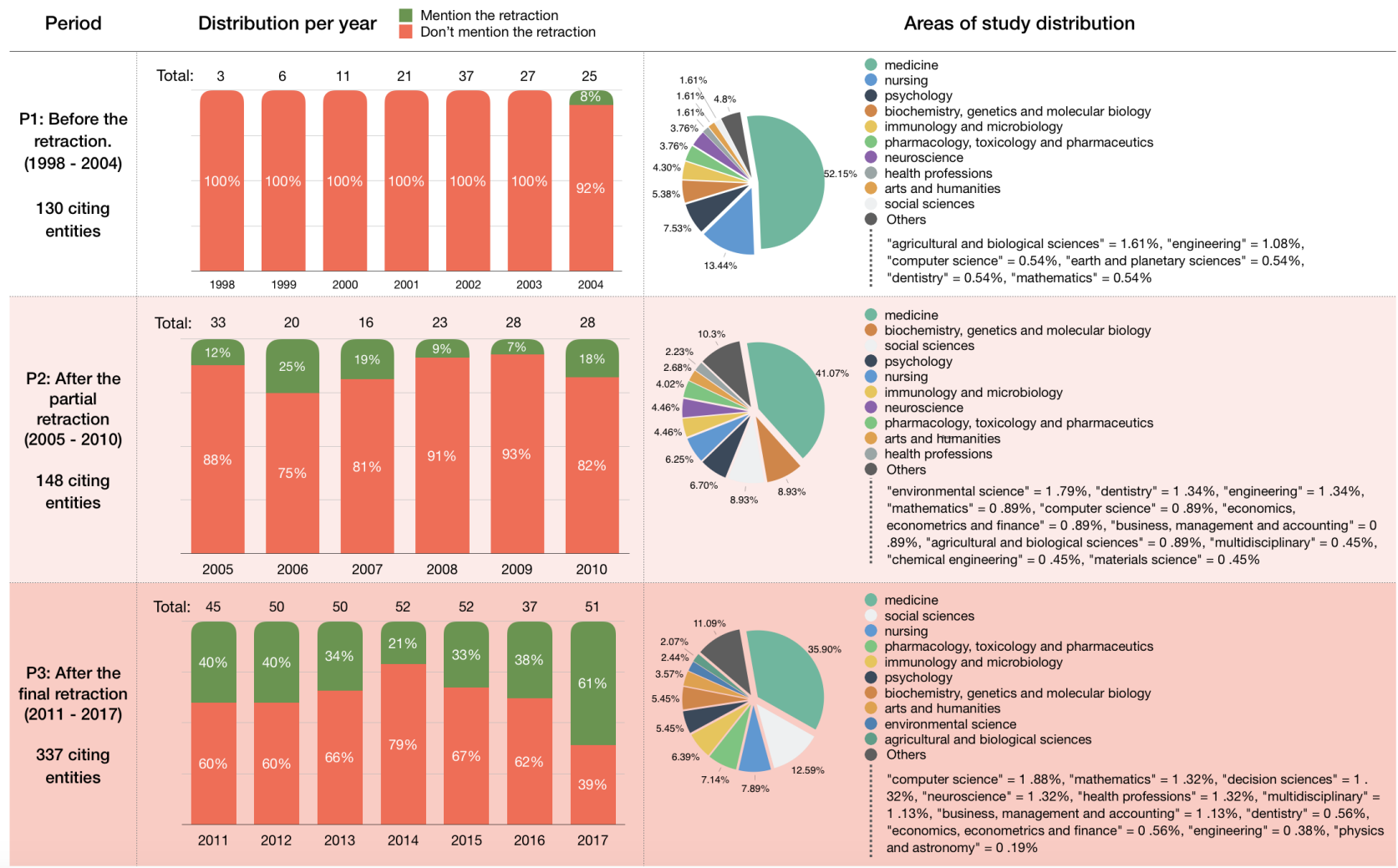

**Figure 4.** A summary of the citing entities. The first column contains the periods P1-P3 we considered, the second column shows the distribution per year of the citing entities that do mention (in green) or do not mention (in red) the retraction of WF-PUB-1998, while the third column shows the distribution of the subject areas of the citing entities.

As anticipated in the previous section, we obtained the topic model using the abstracts of all the publications considered and summarized in Figure 2. Considering the results of the coherence score introduced in Section "Topic Modeling", we built a topic model of 13 topics. Figure 5 shows the related LDAvis visualization. The left part of it shows two different clusters, and one of the clusters is composed of one big topic, i.e. topic 3, which was by far the larger topic identified by the process. Looking at the 30 most salient terms, the term "retract" is in the $5^{th}$ position, meaning that some of the citing entities talked about the retraction of WF-PUB-1998 or, more generally, the retraction phenomena. The same list includes terms such as "social", "movement", "debat", "media" and "cultur" which seem not to be strictly related with medical jargon. This scenario may be an indicator that some of the citing entities are not medical publications. Finally, among these 30 most salient terms, we found terms with a strong negative connotation, such as "fraud".

Using the data obtained through the topic model, we were able to explore each individual topic and give a possible interpretation to it by analyzing its 30 most probable terms, as shown in Table 5 (in Appendix).

The MTMvis visualizations are plotted considering the period P1-P3 (Figure 6) and the subject areas of the citing articles (Figure 7). As shown in Figure 6, the topics 1, 2, and 5 were constantly increasing their percentages over the time while, on the contrary, topics 4 and 9 were decreasing. Along the same lines, topics 3 and 11 showed a very similar pattern along the three periods. As shown in Figure 7, some subject areas, such as *medicine* and *social sciences*, referred to almost all the topics while others (e.g. *computer science*) referred to particular subset of topics.

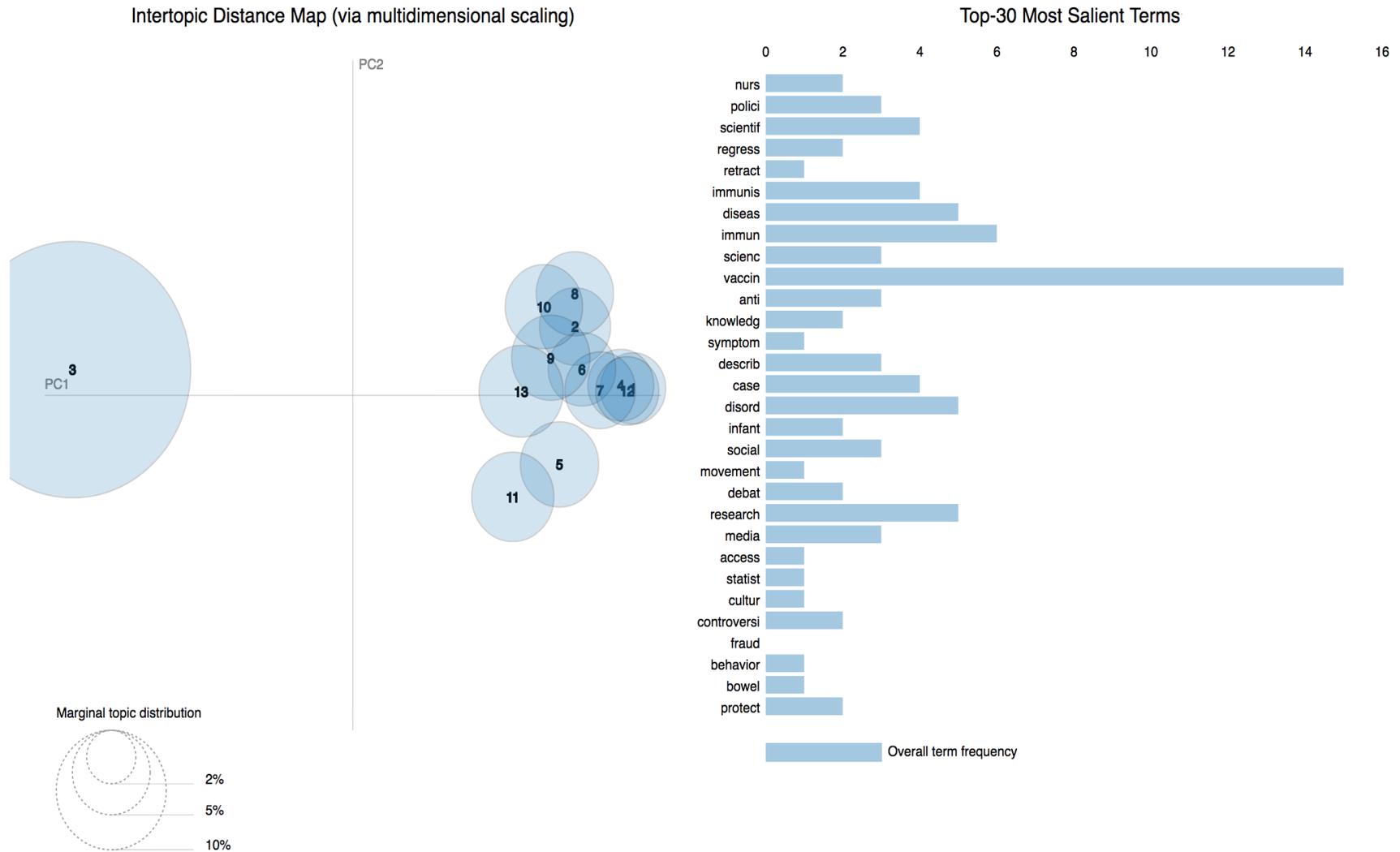

**Figure 5.** The LDAvis visualization built over the topic model obtained from the abstracts of the citing entities.

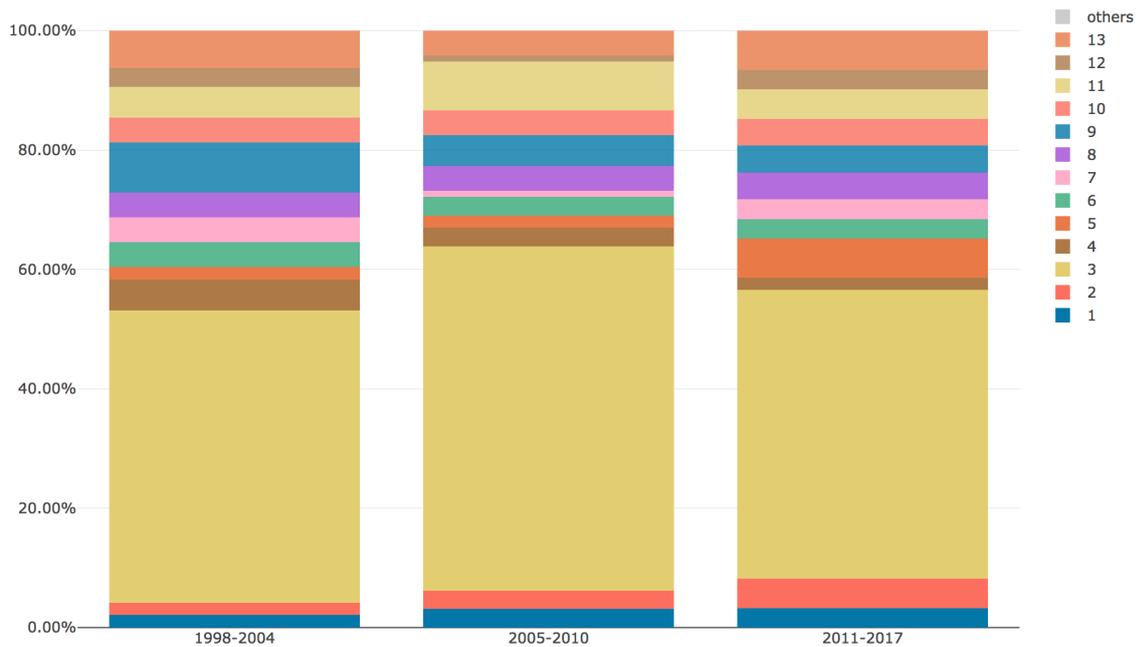

**Figure 6.** MTMvis built on the topic model obtained from the abstracts of the citing entities, shown against the three period P1-P3. For each period the visualization plots the topics distribution (e.g. topic 3 is the dominant topic in all the periods: P1, P2, and P3

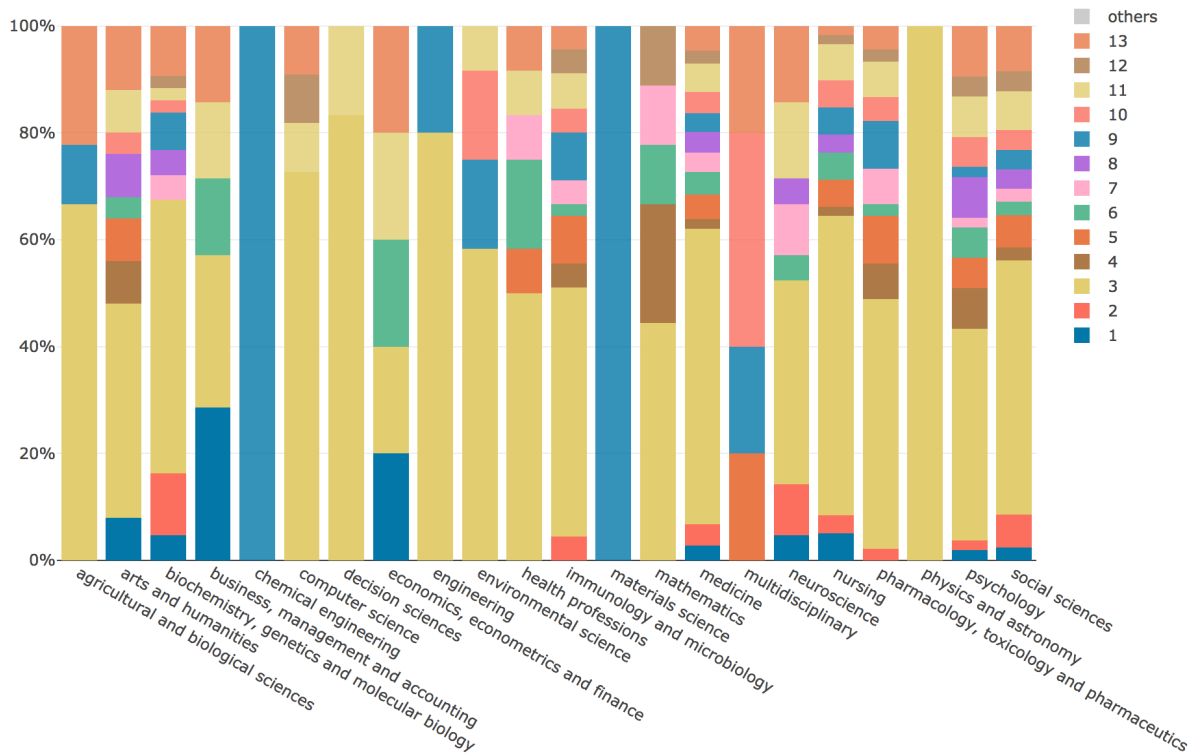

**Figure 7.** MTMvis built on the topic model obtained from the abstracts of the citing entities, shown against their subject areas. For each subject area the visualization plots the topics distribution (e.g. topic 3 is the dominant topic in "arts and humanities")

## In-text citations

The total number of in-text citations to WF-PUB-1998 gathered from the 615 citing entities was 870 (1.4 in-text citations per citing entity on average). In Table 3, we list the features we collected, accompanying them with a brief description (first column) and the corresponding values (second column), i.e. the total number of in-text citations having a value specified for the corresponding feature and, if applicable, a classification of the different possible values.

**Table 3.** The features that directly characterize the in-text citations. The first column lists the features with a brief description, while the second column summarizes the related values we gathered, i.e. the total number and, if applicable, a classification of the different values

| WF-PUB-1998 in-text citations features | Values |
| --- | --- |
| *intext_citation.section*<br>The kind of section in the citing entity which includes the in-text citation, taken from the list in (Suppe, 1998) | **Total:** 757 (87%) in-text citations had a value specified.<br>**Values:** 10 different values: *introduction* (166), *discussion* (61), *results* (28), *background* (36), *conclusions* (17), *method* (15), *abstract* (5) |
| *intext_citation.context*<br>The textual context in the citing entity which includes the in-text citation | **Total:** all the in-text citations had a value specified. |
| *intext_citation.pointer*<br>The string representing the in-text reference pointer (e.g., "[3]") in the citing entity to the bibliographic reference of WF-PUB-1998 | **Total:** all the in-text citations had a value specified. |
| *intext_citation.intent*<br>The citation intent related to the in-text citation in the citing entity, i.e., the author's reason for citing WF-PUB-1998, taken among the citation functions defined in CiTO | **Total:** all the in-text citations had a value specified.<br>**Values:** 17 different values: *discusses* (226), *disputes* (114), *credits* (95), *cites for information* (90), *cites as evidence* (74), *qualifies* (70), *describes* (60), *obtains background from* (56), *critiques* (55), *includes excerpt from* (8), *obtains support from* (6), *uses data from* (5), *uses conclusions from* (4), *ridicules* (4), *extends* (1), *updates* (1), *refutes* (1) |
| *intext_citation.sentiment*<br>The sentiment, classified as positive/negative/neutral, conveyed by the citation context of an in-text citation | **Total:** All the in-text citations had a value specified.<br>**Values:** *neutral* (549), *negative* (300), *positive* (21) |

Figure 8 shows descriptive statistics of some of the values introduced in Table 3. The sentiment is combined with all the statistics displayed (red for negative, yellow for neutral, green for positive). The first column contains the three periods P1-P3 considered in our analysis. The second column shows the distribution per year of the in-text citations, the third column shows the distribution of citation intents, and the fourth column shows the distribution of the sections where in-text citations were contained. The sections are classified considering the list proposed in (Suppe, 1998) when possible, while all the others are grouped under the label *"Others"* (i.e. a section with a generic title which could not be identified in any section from the proposed list in (Suppe, 1998)).

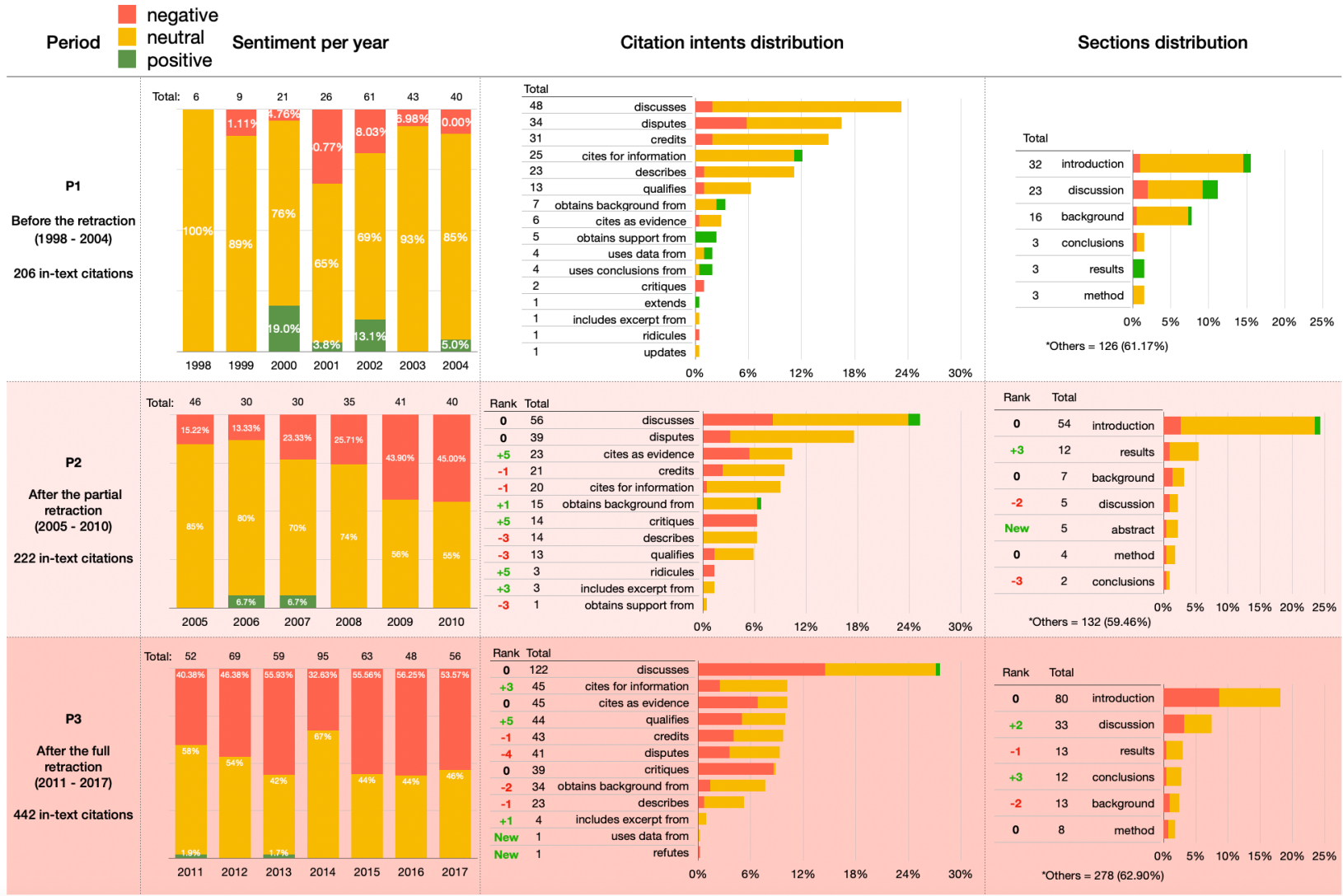

**Figure 8.** A summary of the in-text citations. All the data are classified under the three sentiments: negative (red), neutral (yellow), and positive (green). The first column contains the periods P1-P3 we considered, the second column shows the distribution per year of the in-text citations, the third column shows the citation intents distribution, and the last column shows the in-text citation sections distribution.

Figure 9 shows the LDAvis of the 22 topics we retrieved using the topic modeling methods described in Section "Topic modeling" by using the citation contexts of in-text citations to WF-PUB-1998. In contrast with the analysis conducted on the abstracts of the citing entities, the 30 most salient terms did not include any term related with the retraction phenomena. The sparsity of the topics in this LDAvis is higher than the one observed with the abstracts and allowed us to spot three different clusters. In particular, we observed two topics with a high prevalence which are also very distant among them (topics 8 and 12). Table 6 (in Appendix) lists all the topics and provides our own interpretation according to their 30 most probable terms.

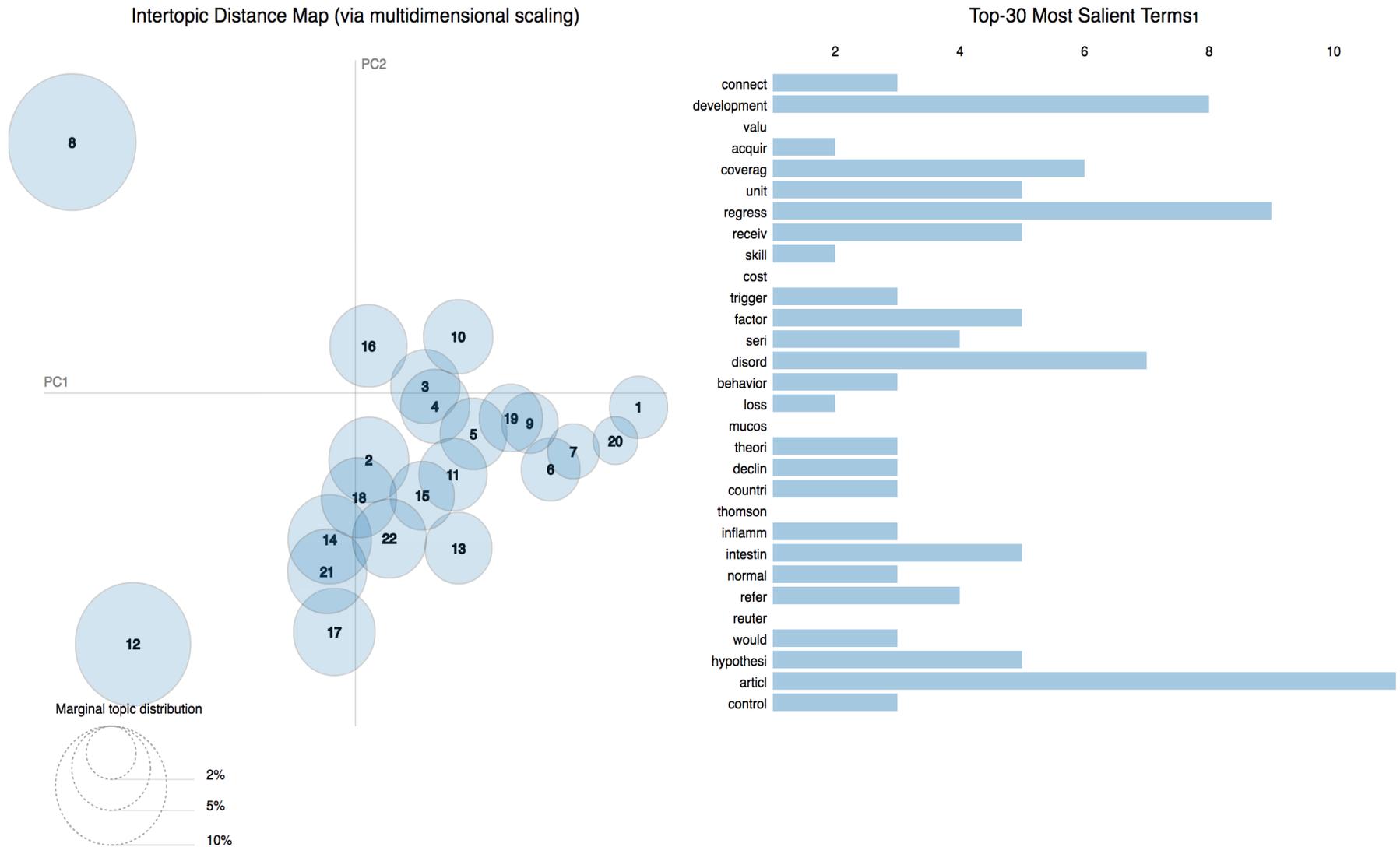

**Figure 9.** The LDAvis visualization of the topic model created using the citation contexts of the in-text citations contained in the entities citing *WF-PUB-1998*.

The MTMvis visualizations in Figure 10 and Figure 11 refer again to the distribution of the topics over P1-P3 and considering the subject areas of the entities containing the in-text citations analyzed. Figure 10 shows that topics 1, 5, 6, 12, and 22 were constantly increasing their percentages throughout P1-P3. Topic 8 and 16, instead, were significantly decreasing along the same period. Topics 2, 4, 10, 13, 19, and 21 had a similar behavior across P1-P3, having their peak in P2. Contrarily, topics 3, 9, 14, and 18 showed a clear decrease in P2, while in P1 and P3 they showed a similar (and higher) presence.

Figure 11 shows that *medicine*, *social sciences* and *nursing* were the areas of study that included the larger part of the topics identified. In addition, we also had subject areas with a high number of topics which do not concern the medical and social science domains, i.e. *agricultural and biological sciences*, *arts and humanities* and *computer science*.

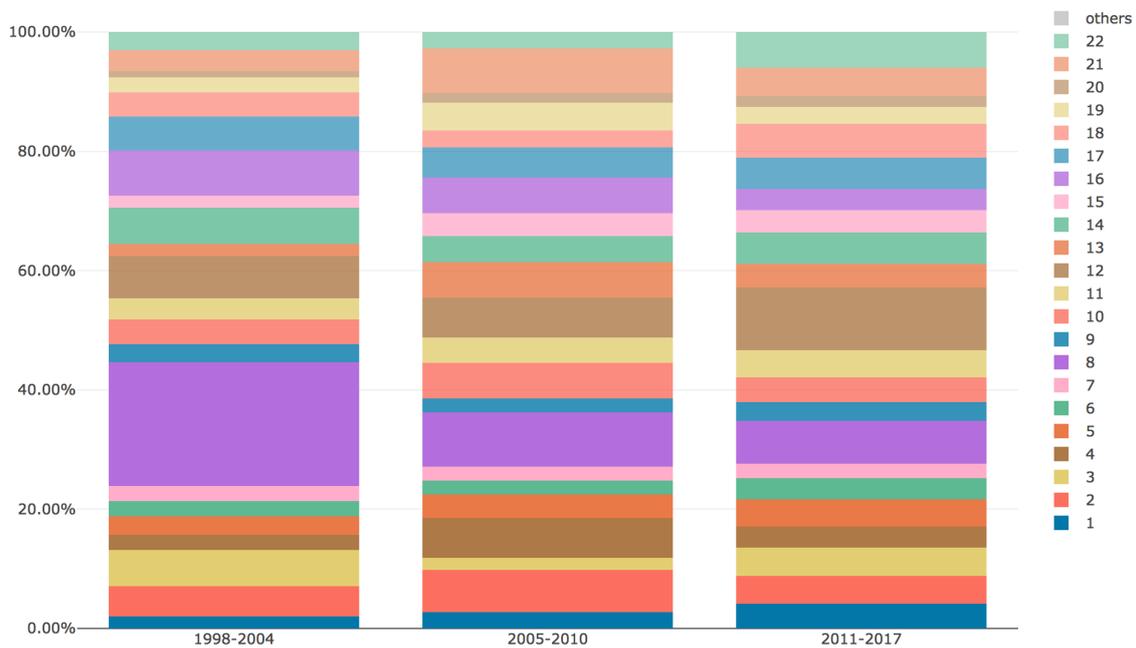

**Figure 10.** MTMvis created considering the topics extracted from the citation contexts of the in-text citations citing WF-PUB-1998 according to the periods P1-P3. For each period the visualization plots the topics distribution – e.g., topic 8 (in purple) is the dominant topic in P1.

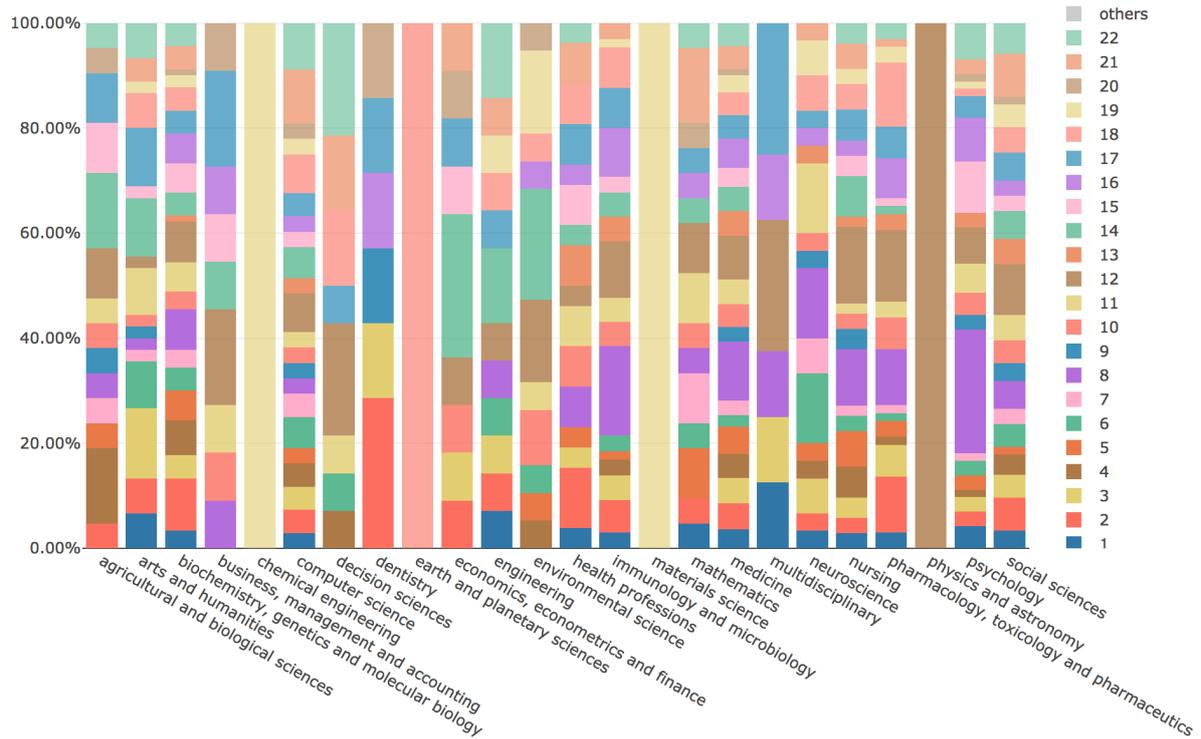

**Figure 11.** MTMvis created considering the topics extracted from the citation contexts of the in-text citations citing *WF-PUB-1998* according to the subject areas of the citing entities. For each period the visualization plots the topics distribution – e.g., topic 3 (in dark yellow) is the dominant topic of the "arts and humanities" subject area.

# Discussion

In this section, we discuss the result introduced previously, and we provide insights to answer the two research questions presented in Section "Introduction". Also, we introduce some limitations of our study and provide suggestions on how to address them in future investigations.

## Answering RQ1

In this section we address RQ1: what are the research topics introduced in the articles citing WF-PUB-1998 before and after its retraction?

From a quantitative point of view, while looking at the subject areas of the citing entities we gathered (see Figure 4), we noticed an increment in the number of areas involved in time. Indeed, the total number of subject areas were 17 in P1 (i.e. before the first partial retraction), while in P2-P3 we counted 22 different subject areas. In addition, in P2-P3 we observed a higher prevalence of non-medical subject areas. Considering the percentage value in P3 with respect to the one in P1, then *social sciences* and *arts and humanities* had increased their percentages, respectively, of 7.81 and 2.21 times more than those observed in P1. On the contrary, considering the same periods (P1 and P3), *medicine* and *nursing* had an inverse trend, since their presence decreased by almost 30% and 40% percent compared with P1, respectively. These figures suggested that the retraction attracted the attention of other subject areas which were not strictly related to the original one of WF-PUB-1998 (i.e. *medicine*).

In addition, we also noticed a continuous increment in the percentage of entities that have explicitly mentioned the retraction of WF-PUB-1998 over the time (see Figure 4). The peak was reached in 2017 (the last year we have considered) with a 61% percentage of entities mentioning the retraction, and we had an important percentage of entities mentioning the retraction even before the full retraction notice (e.g. 25% of entities in 2006). This suggests that the citing entities do not always wait for the full retraction notice before acknowledging the retraction (even if partial). This aspect might be also related to the particular kind of the partial retraction (that was "Concerns/Issues About Results", and "Error in Results and/or Conclusions" in WF-PUB-1998) and with the popularity of the particular case in consideration.

Looking at the retrieved topics in the topic model created using the abstracts of the citing entities, we noticed that topics 1, 2 and 5 were those increasing their presence after the partial retraction (i.e. starting from P2). The themes covered by these topics seemed to refer to discussions on the retraction phenomena (see Table 5 in Appendix) and used a limited number of terms from medical jargon.

A deeper investigation of the evolution of topics 1, 2 and 5 during P2-P3 on all the subject areas, showed that topics 2 and 5 had got a significant increment in P3 (11.48% vs 5.15%) while topic 1 has a slighter increment (3.09% vs. 3.28%), as we can see in Figure 12. This might indicate that topic 1 (and the abstracts linked to it) discussed the retraction phenomena similarly over P2-P3. In fact, although topic 1 included words that deal with ethical/social issues (see Table 5 in Appendix), it did not include words strongly related to the retraction or having a strongly negative sentiment. The citing entities linked to topic 1 cited WF-PUB-1998 and discuss the case without mentioning the actual retraction of WF-PUB-1998, even after its full retraction (i.e. P3). Figure 13 shows that topic 1 is mainly related to the *medicine* subject area (excluding the subject areas with limited number of abstracts, e.g. *arts and humanities* with 2 abstracts). This relation between topic 1 and *medicine* is also interesting: indeed, topic 1 has little engagement with the medical themes, considering its 30 most probable terms. Thus, part of the entities in the *medicine* subject area discussed the retraction of WF-PUB-1998 in non-medical terms as well.

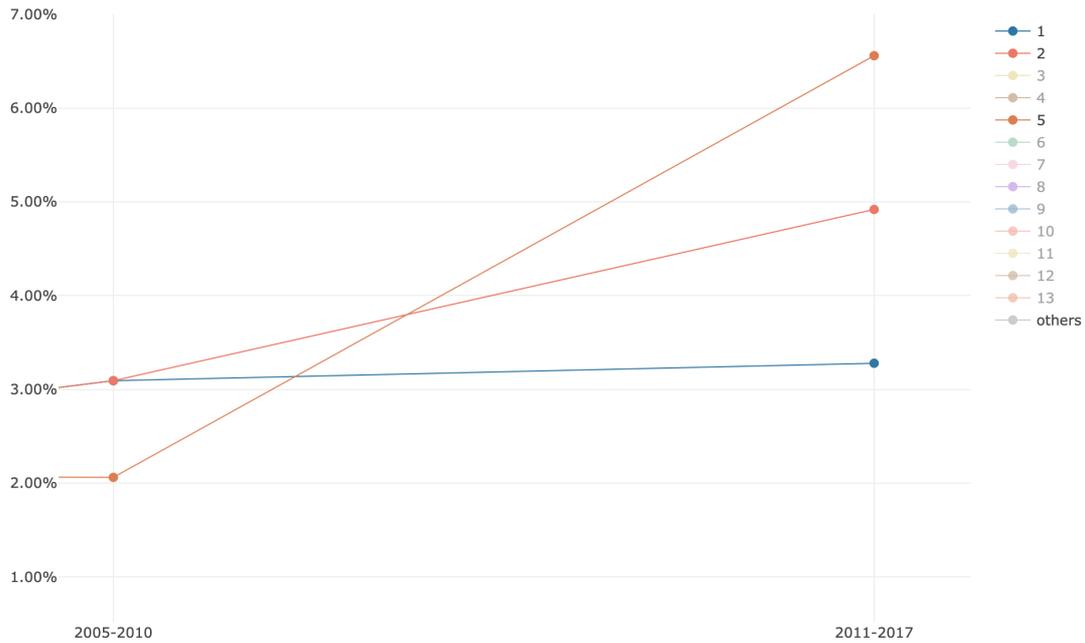

**Figure 12.** The evolution of topics 1, 2 and 5 during P2-P3 on all the subject areas plotted using MTMvis. MTMvis has been generated from the topic model created using the abstracts of the citing entities. The themes covered by these topics are close to the retraction phenomena and used a limited number of terms from medical jargon.

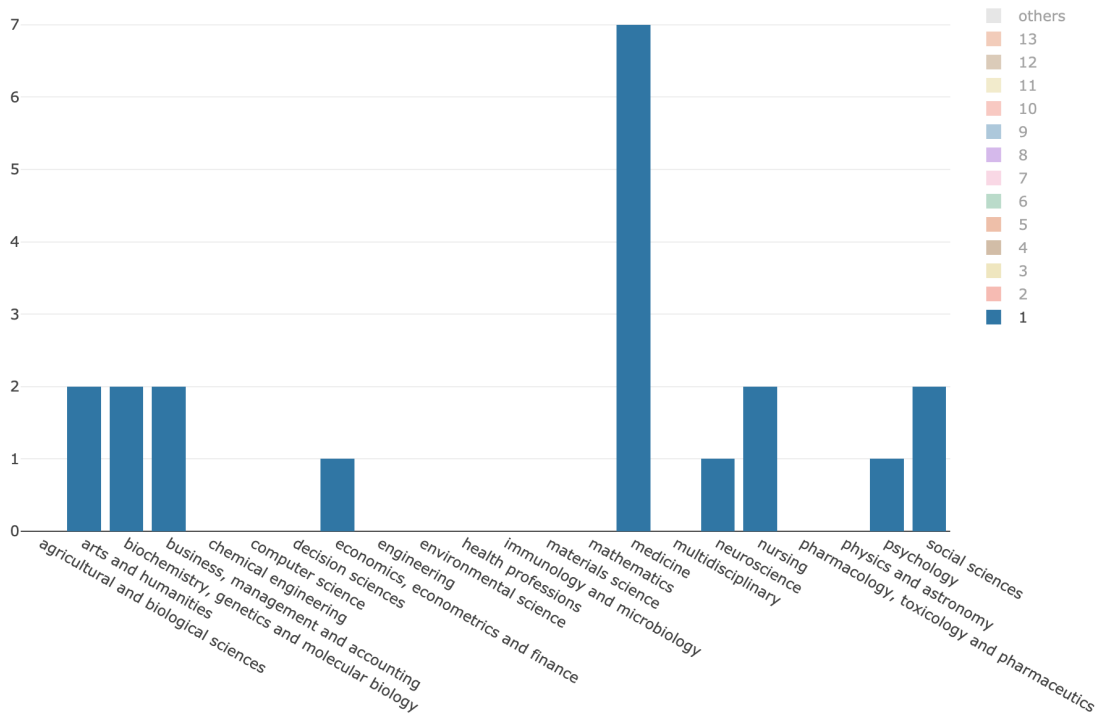

**Figure 13.** The distribution of topic 1 over all the subject areas during P2-P3 plotted using MTMvis. MTMvis has been generated from the topic model created using the abstracts of the citing entities. Topic 1 include terms from the social science domain and relates to ethical themes.

We investigated the distribution of topics 2 and 5 over the subject areas during P2-P3 and checked if such topics were part of the top five ones of each related subject area, as summarized in Figure 14. We can see that topics 2 and 5 were listed in the top five topics of twelve subject areas. Avoiding considering the subject areas for which we had a small number of abstracts in P2-P3 (e.g. *economics, econometrics and finance* and *multidisciplinary*, both having 1 abstract), we noticed that topics 2 and 5 were highly represented in the *social sciences* subject area with a total percentage of 12% (number of abstract: 12) of all the abstracts in P2-P3. These considerations suggest that topics 2 and 5 were the ones that better represent and characterize the period after the full retraction (i.e. P3), and that *social sciences* is the subject area that dealt the most with the themes emerged in P3. Contrary to our previous considerations regarding topic 1, in these two topics we found a clear reference to the retraction. The fact that this aspect was manifested in the analysis of the abstracts may indicate that the retraction might have been one of the main subjects discussed in the entities of the abstracts analyzed.

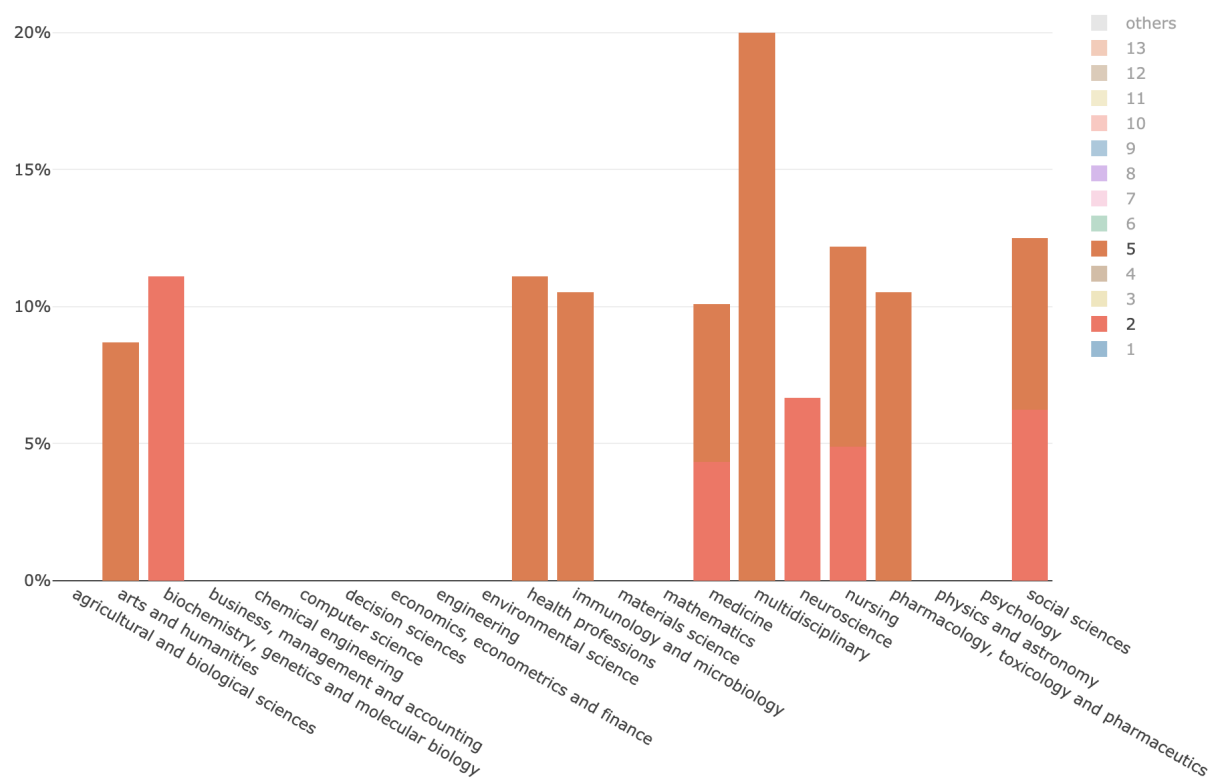

**Figure 14.** The subject areas of citing entities published in P2-P3 which includes either topic 2, or 5 in their top 5 topics. The themes covered by these topics relate to the retraction phenomena and use a limited number of terms from medical jargon.

## Answering RQ2

In this section we address RQ2: what are the most relevant characteristics of the in-text citations (e.g. intent, sentiment, mention of the retraction, etc.) in the articles citing WF-PUB-1998 before and after its retraction?

Figure 8 shows that the intended sentiment carried in the citation contexts of the in-text citations referring to WF-PUB-1998 moved to the negative spectrum over time. However, the retraction of WF-PUB-1998 was not always mentioned in these cases. Indeed, as shown in Figure 8, in 2015 only 32% of the citing entities mentioned the retraction even if the perceived sentiment in the same year is either negative (for 55.56% of in-text citations) or neutral (for 44% of in-text citations).

The distribution of the citation intents annotated in the in-text citations during P1-P3 showed an increment in the use of generic intents such as *discusses* and *cites for information*. This could be related with increasing popularity of the retraction of WF-PUB-1998 in the non-medical subject areas (as already stated in the previous section). Probably, the entities that are part of the non-medical subject areas cited WF-PUB-1998 with a generic intent without recalling strictly medical details in their text, which are out of the scope of their research domains.

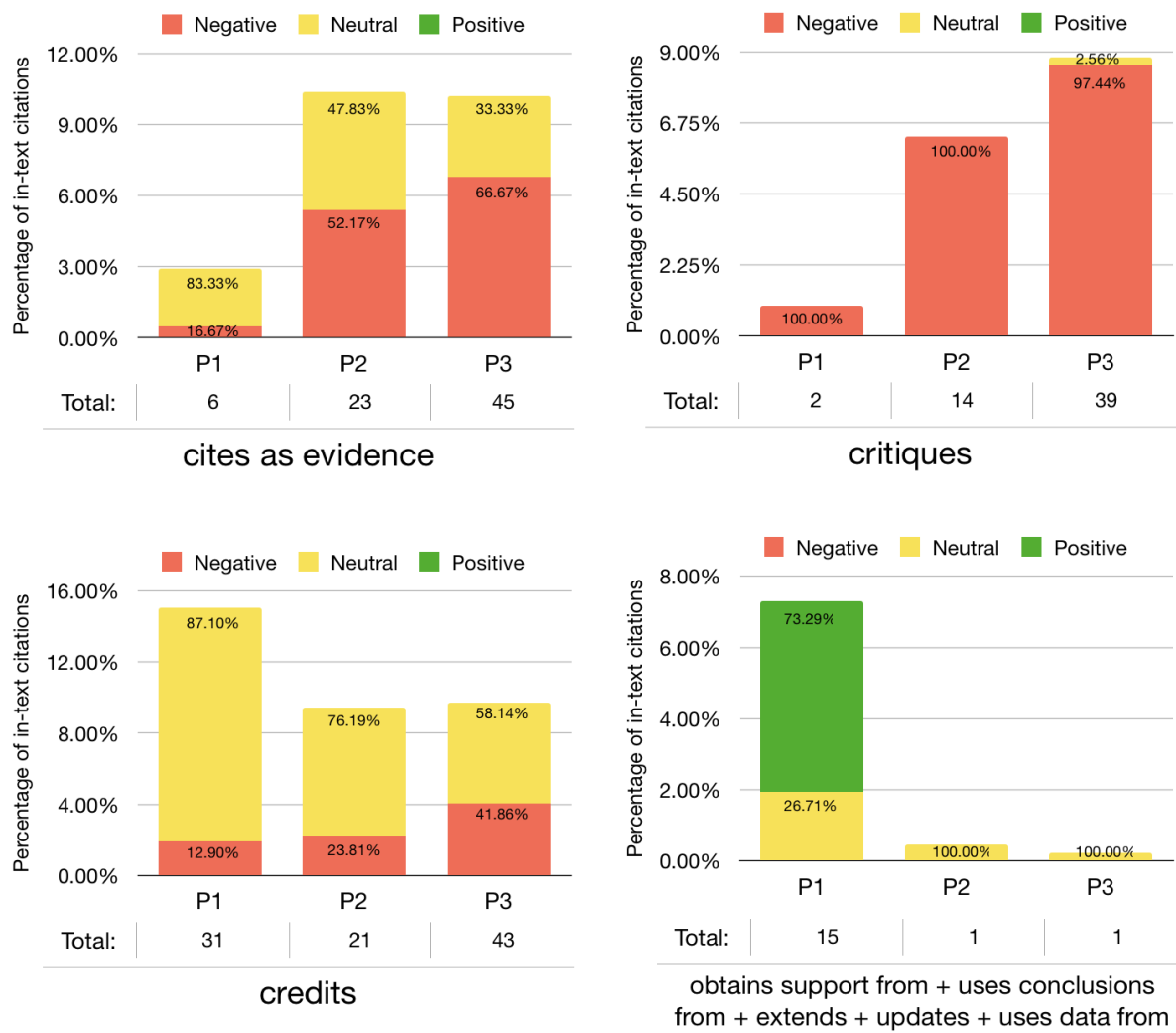

**Figure 15.** The four graphs illustrate the way the use of citation intents changed over time (i.e., the three periods P1, P2, and P3) and according to their perceived sentiment. The citation intents *cites as evidence*, *critiques*, and *credits* are illustrated in separated charts, that show an increment in the negative sentiment along the three periods.

As shown in Figure 15, the set of intents *uses conclusions from*, *updates*, *extends*, *uses data from*, and *obtains support from* decreased starting from P2, probably due to a lesser use of the data and conclusions contained in WF-PUB-1998 after its retraction. Other citation intents, instead, showed a clear increment of their use along the three periods. For instance, the use of *critiques* seemed to be related somehow with the increment of the negative sentiment overall.

Instead, *credits* had an important drop. In this case, the citing entities published before the partial retraction of WF-PUB-1998 used it mostly in a neutral way to credit Wakefield and colleagues for their findings. However, in P2-P3, beside the overall drop, *credits* had a higher percentage of negative citations. This last aspect was also noticed in the intent *cites as evidence*, although its overall usage has increased in time. However, if before the retraction, *cites for evidence* was used neutrally to refer to WF-PUB-1998 to support some statements or conclusions in the citing entities, after the retraction it was actually used to highlight WF-PUB-1998 as a negative scientific example due to its retraction and, more generally, of a faulty science.

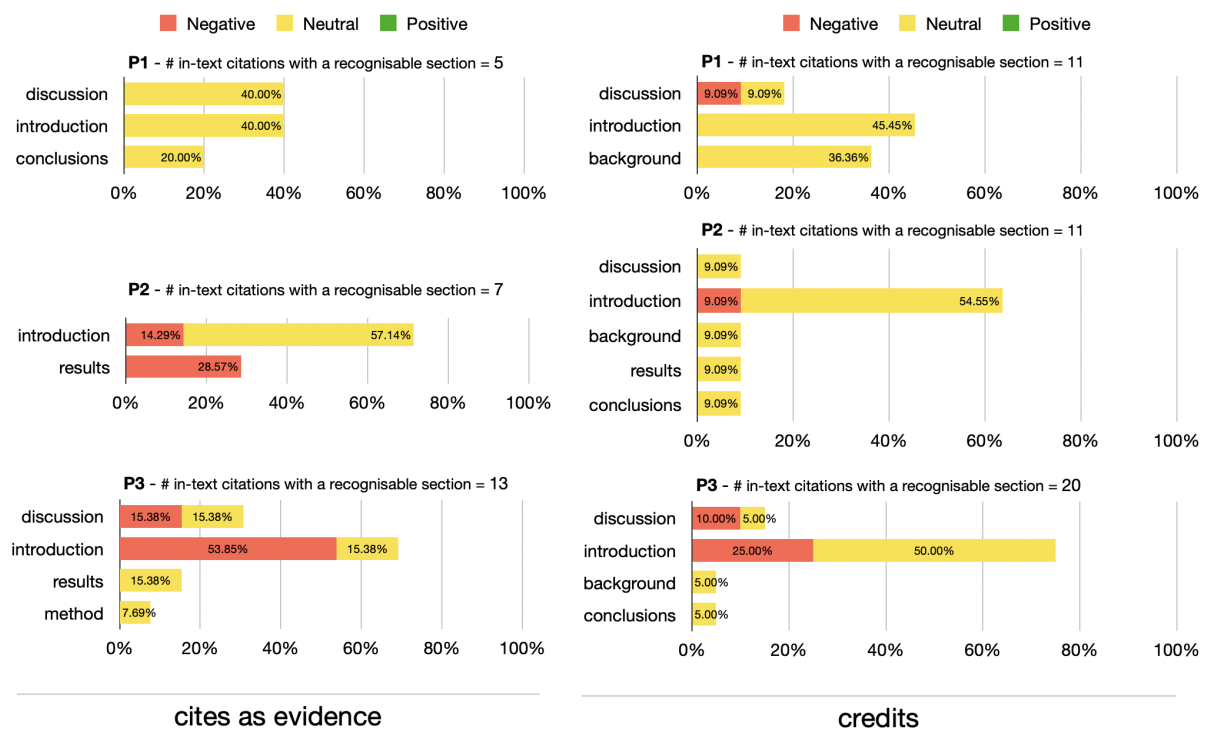

**Figure 16.** The *cites as evidence* and *credits* citation intents distributions among the sections (the recognizable ones) and during the three periods (i.e. P1-P3).

In Figure 16, we investigated the sections of the in-text citations marked as *credits* and *cites as evidence*. On the one hand, the *credits* citations were mostly distributed on descriptive sections – i.e. *introduction, discussion,* and *background* – during all the three periods. On the other hand, the *cites as evidence* citations appeared also in technical sections – i.e. *results,* and *method*. The sections distribution in P3 for both *credits* and *cites as evidence* followed the overall distribution introduced in Figure 8: the in-text citations have been concentrated in few sections mostly of descriptive type – i.e. *introduction* and *discussion*.

We analyzed the twenty-two topics we obtained considering the topic model created using the citation contexts of the in-text citations referring to WF-PUB-1998. In particular, as shown in Figure 17, we focused on:

1. the topics for which we observed an increasing use over time;
2. the topics which had a huge increment in their use in P3;
3. the topics which had a constant decrement in their use over time.

The topics that increased over P1-P3 (i.e. topics 1, 5, and 11) included a few medical terms and seemed to refer to the controversy of the retraction of WF-PUB-1998 from a mathematical and statistical perspective. A second group of topics (i.e. topics 12, 18 and 22) seemed to refer to WF-PUB-1998 as an example of faulty science, which was acknowledged clearly in P3. The drastic change of these topics in P3 is very significant. Indeed, all the three topics (as shown Table 6 in Appendix) mention the word "retraction" (and its derivatives) along with other words with a strong negative connotation. In other words, it seems that the authors waited the full retraction notice before marking their negative impressions toward WF-PUB-1998 – 19.8% of the citations in P3 are part of this group of topics.

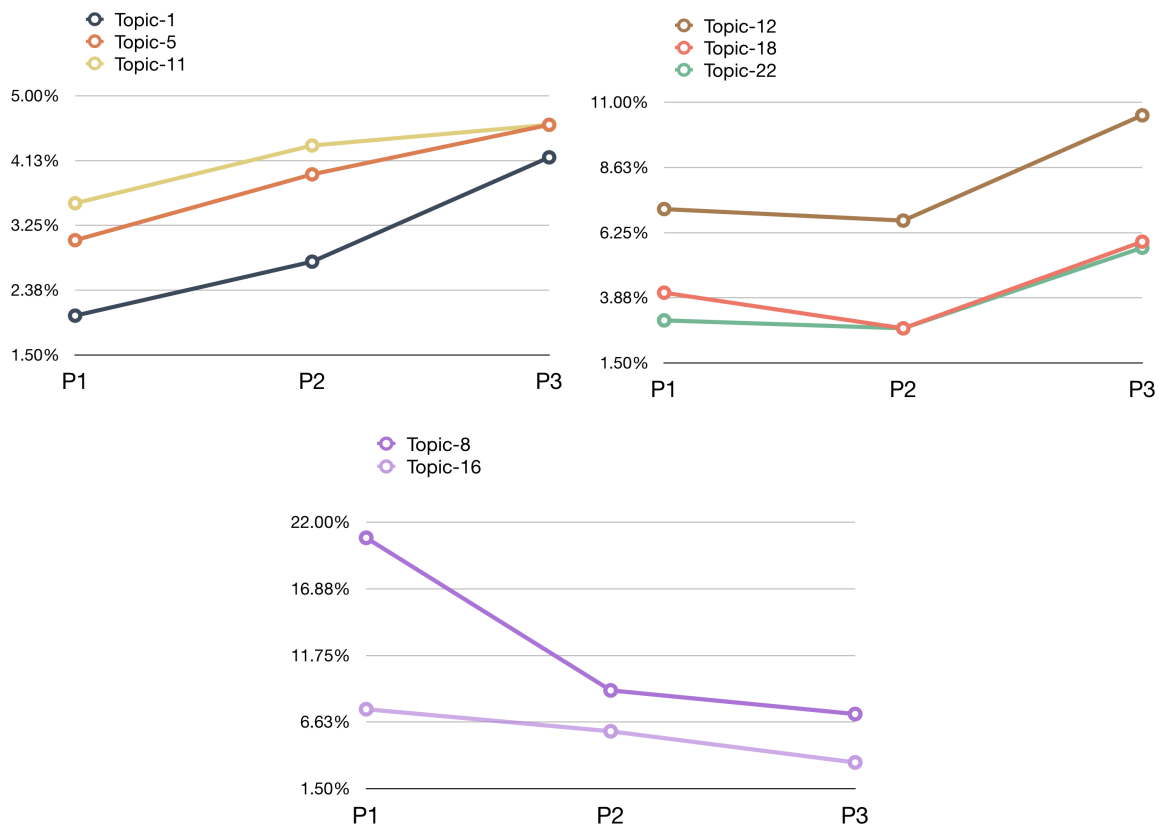

**Figure 17.** The evolution over time of three groups of topics defined from the citation contexts of the in-text citations to WF-PUB-1998.

Similar behavior could be noticed also in the citations coming from medical subject areas, since 22.97% and 30.61% of the citations in P3 are coming from *medicine* and *nursing* articles, respectively. This suggests that also the entities close to the domain of the retracted article did not hesitate to judge a retracted work done by their colleagues.

The last group of topics (i.e. topics 8 and 16) were mainly related to the medical domain, and included some medical themes treated in WF-PUB-1998. The fact that these topics had a clear decrement over time suggests that the most recent citing works provided partial and limited acknowledgement of the conclusions and medical arguments in WF-PUB-1998.

In Figure 18, we show the topics that either increased (left panel) or decreased (right panel) their presence over time considering only the citation contexts of the citing entities belonging to the *medicine* subject area. Some of the topics shown in Figure 18 are also included in Figure 17, although there is an important difference: topic 15 (that concerned the conclusions of WF-PUB-1998 and the controversies arising from it) is not listed in Figure 17, even if it seemed relevant when we focus only on the *medicine* subject area. We had a similar situation also with the topics decreasing over time. Indeed, topic 7 (which summarizes WF-PUB-1998 and medical conclusions) is not highlighted in Figure 17 as well.

This scenario suggests that the citing entities in the *medicine* subject area included additional prominent topics when discussing WF-PUB-1998. More precisely, after its final retraction (i.e. P3), part of the entities addressed the retraction through a discussion using medical terms. The decreasing relevance of topic 7 indicates that also the entities part of the same subject area of the retracted article (i.e. *medicine*) addressed less the subjects treated in WF-PUB-1998, and rather focused on citing and discussing the retraction of WF-PUB-1998 without deepening its medical themes.

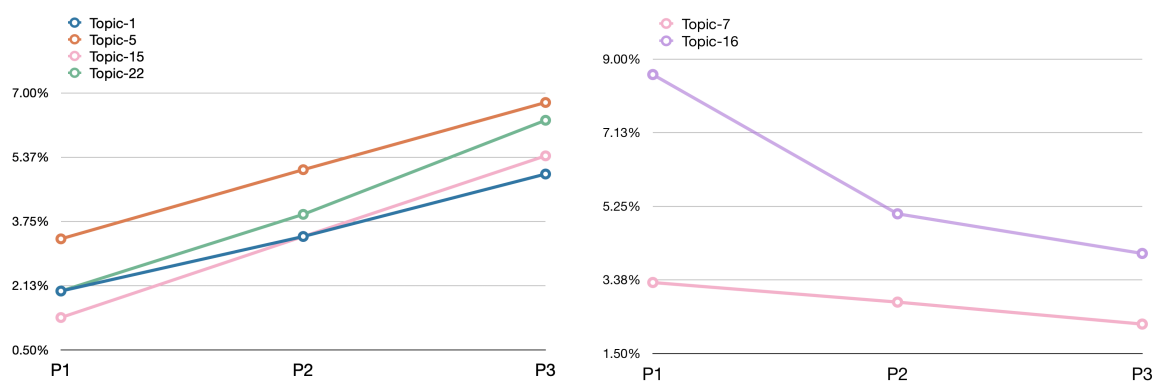

**Figure 18.** The increasing (left) and decreasing (right) topics of the in-text citation topic model, considering only the *medicine* area of study.

## Limitations of our study and future suggestions

Our findings and observations provide additional insights on the retraction of WF-PUB-1998 and how it has been perceived by the scientific community. However, we are aware of particular limitations that may have affected the findings and the interpretations we made throughout this study. In this section we list the methodological limitations, and we compare our outcomes with previous works on the same topic.

First, we used the data in COCI to gather all the citations to WF-PUB-1998 used in our study. Since COCI contains citations between entities included in Crossref when they are both identified by DOIs, we did not include in our analysis all the citations that involved entities with no DOIs. Also, we missed the citations to WF-PUB-1998 from articles published by some publishers, such as Elsevier, that did not share openly their reference lists via Crossref in 2018 – and that, thus, were not available in the COCI dump, i.e. the November 2018 release (OpenCitations, 2018).

For a few citing entities (i.e., 22) involved in the citations we gathered, we could not retrieve their full text due to commercial paywalls, preventing us to analyze the citation contexts and in-text citations they defined. Thus, we excluded these citing entities, and their related citations, from our analysis.

**Table 4.** A summary of the differences and similarities between our study and (Suelzer *et al.*, 2019).

| | Method and results in (Suelzer *et al.*, 2019) compared with our study | |
|---|---|---|
| **Feature** | **Method** | **Results** |
| **Analyzed entities and the citation context** | **Differences:** <br> 1) The analysis does not address and collect the in-text citations as separated entities. <br> 2) The reviewed in-text citation contexts are not defined following a precise structure/rule – in our work, indeed, the context is defined by the anchor sentence, plus the prior and subsequent sentence. <br><br> **Similarities:** <br> 1) the citing works are collected and analyzed as individual entities. <br> 2) the definition of three periods following the years of the partial and full retraction (i.e. P1, P2, and P3). | **Differences:** <br> 1) The total number of citing entities is higher – 1,153 (compared to the 615 of our work). <br> 2) The analysis does not mention any statistics regarding the number of in-text citations. <br><br> **Similarities:** <br> 1) The percentage of citing entities gathered starting from the partial retraction to the last date considered (i.e. March 11, 2019) with respect to the total number is 76% (close to the 78% of our work). Generally, considering the periods P1, P2, and P3 the distribution is respectively 23%, 28%, and 49% (close to our distribution: 22%, 24%, and 54%). |
| **Mentions of the retraction** | **Differences:** <br> 1) There is no definition of the textual contexts gathered in order to establish whether the citing work had mentioned the retraction or not. <br> 2) It also checks whether the retraction is mentioned in the reference list. <br><br> **Similarities:** <br> 1) A citing entity mentions the retraction, only if the word "retraction" (and its derivatives) is used. | **Differences:** <br> 1) A higher percentage of entities mention the retraction after the partial retraction – i.e. P2-P3 (56% vs 25%). This gap is reduced when considering only the retraction mentioned in the citation context – 38.3% vs 25%. <br><br> **Similarities:** <br> 1) The overall trend in P2-P3 is similar and shows a continuous increment in the number of entities mentioning the retraction over time. <br> 2) 2009 was the year with the lowest number of entities mentioning the retraction. |
| **Citation Intent** | **Differences:** <br> 1) The citations are characterized into 8 different categories, following the definitions in (Bornmann & Daniel, 2008). <br> 2) There is no distinction between the intent and the sentiment of the citation (e.g. a citation could be characterized as *methodologic* or *negative*, but there is no way to combine both these dimensions). <br> 3) The citation characteristics address the citing works entities (on our work the annotation refer to the specific in-text citations). <br><br> **Similarities:** <br> 1) The annotation was performed following a set of rules which guided and helped the annotator. | **Differences:** <br> 1) The plotted results combine in the same dimension both the sentiment and the citation function, following the established annotation rules, thus this fact makes the comparison with our results difficult. |
| **Citation sentiment** | **Differences:** <br> 1) It is not part of the analyzed features, although this subject is addressed in the discussion section. <br> 2) This information is embedded in the citation intent value (e.g. using the characteristics *negative*, or *affirmative*) | |
| **Citation section** | **Differences:** <br> 1) It is not part of the analyzed features, although this subject is part of the discussion since some articles are provided as examples of citations appearing in the *introduction* and *discussion* sections. | |
| **Text analysis (i.e. topic modeling)** | **Differences:** <br> 1) It is not part of the analyzed features. They give a brief summary regarding their observations of some examples that have cited WF-PUB-1998. | |

While working on similar problems, the data we gathered in our study are slightly different from those used in (Suelzer *et al.*, 2019), which introduces an analysis of WF-PUB-1998, as anticipated in the introduction. In particular, Suelzer *et al.* collected 1,211 articles gathered from the Web of Science Core Collection in March 2019, while we collected citations coming from 615 articles in total up till November 2018. This number disparity is strictly related to the prior date, along with the fact that we relied only open citations repositories (COCI in particular) in order to foster the reproducibility of our analysis. We are aware that this might have an impact regarding the final results, although we believe that using openly available resources strengthens the applicability of the methodology.

The gap between open and non-open citations should be significantly reduced in the next releases of COCI, due to (a) the recent decision of Elsevier of making reference lists of all articles openly available via Crossref (https://www.elsevier.com/connect/advancing-responsible-research-assessment), and (b) the Crowdsourcing Open Citations Index (CROCI) (Heibi *et al.*, 2019a) which enables scholars and publishers to provide their own citations to OpenCitation to upload them into the OpenCitations Indexes. In principle, these activities will increase the number of citations that can be gathered using our proposed methodology. Other similarities and differences between our study and Suelzer *et al.*'s one are introduced in Table 4.

Another important aspect of our study is the manual annotation of citation intents. Although the annotation has been done carefully by following a specific methodology, it was based on a subjective interpretation of the text and, thus, may differ from the original citation intent that the authors of the citations to WF-PUB-1998 had in mind.

In addition to the limits regarding our methods and findings, there are also other aspects that this work did not address compared with the past approaches. In particular, we would like to work, in future developments of this research, in the generation of a citation network starting from either our seed retracted article or from its citing entities, as suggested by van der Vet and Nijveen (2016) who proved the importance of such analysis, since it might enlighten us on the negative/positive outcomes of the propagation of retracted research results.

# Conclusions

In this article, we have presented the outcomes of a citation analysis of a highly cited and popular retracted article: WF-PUB-1998 (Wakefield *et al.*, 1998). We have applied a quantitative and qualitative analysis of the citations that cited WF-PUB-1998 and we clustered them into three periods: (P1) before the WF-PUB-1998 partial retraction, (P2) after the partial retraction and before its full retraction, and (P3) after its full retraction. The main purpose of this work was to understand the retraction phenomenon and how it was perceived from the scientific community when referring to retracted articles in their own work. WF-PUB-1998 is a popular example of a retracted article that was highly cited by other works overtime (before and after the retraction notes), therefore we considered it as a perfect example to analyze. We approached our general goal through the definition of two research questions aiming at analyzing possible evolution, before and after the retraction, of the research topics addressed by the articles citing WF-PUB-1998 and the main characteristics of such citations. To answer these questions, we have defined a methodology which allowed us to gather data, to automatically process the textual information retrieved (abstract and citation context) to extract topics (using a topic modelling technique) and, thus, to address the research questions.

Our results have been presented according to the entities we analyzed: entities citing WF-PUB-1998 and their in-text citations. We first showed a quantitative overview of the features we have collected, and then we discussed the outcomes of the topic models obtained. Finally, in Section "Discussion", we discussed all the evidence we have collected to answer the research questions. In particular, we observed that:

a) the citing entities generally did not wait for a full retraction notice before acknowledging the retraction of the cited article;
b) the *social sciences* subject area is the one that dealt the most with the retraction of WF-PUB-1998;
c) the authors of the citing articles introduce WF-PUB-1998, after its retraction, from a general perspective without recalling strictly medical details in their text.

Finally, we have also discussed the limits of our approach from a methodological point of view, and we compared our methods and results with the ones in (Suelzer *et al.*, 2019). The bigger difference has regarded the additional features we have considered in our analysis – i.e. the citation sentiment, the citation section, and the topic modeling analysis. Many of our findings have also confirmed the results of Suelzer *et al.*'s work.

# Appendix

**Table 5.** The 13 topics available in the topic model obtained from the abstracts of the articles citing WF-PUB-1998. For each topic (row) we mention its proportion percentage in the corpus (column 1), and the 30 most relevant terms (column 2), and we give our interpretation of it (column 3).

| Topic (proportion) | Terms (the 30 most probable terms) | Interpretation |
|---|---|---|
| 1 (3.5%) | *bias, epidem, philosoph, cell, experi, consum, behavior, protect, even, illich, acetaminophen, scientist, scienc, oxid, call, conserv, long, public, scientif, phone, campaign, occur, experiment, ethic, feed, among, politician, reject, dissent, insignific* | Close to the social studies domain, might take in consideration ethical thematic. |
| 2 (4.2%) | *retract, symptom, disabl, student, scienc, perspect, expertis, misconduct, forens, paper, probabl, studi, mandat, cite, diseas, treat, provid, research, replic, vaccin, tripl, librari, qualif, adjuv, might, take, younger, case, media, wakefield* | Includes terms related to the retraction phenomena (study-domain independent), it includes terms used in scientometric analysis. |
| 3 (46%) | *vaccin, parent, children, health, inform, immun, public, decis, review, safeti, studi, measl, evid, articl, research, risk, disord, practic, autist, issu, factor, concern, diseas, report, increas, import, relat, child, effect, base* | This is by far the largest topic out of the 13, it contains common terms, highly frequent in the corpus, and close to the *WF-PUB-1998* thematics. |
| 4 (3.6%) | *access, nurs, bowel, knowledg, hepat, polici, global, mobil, portfolio, immunis, ask, biblic, newspap, pharmaceut, huge, visitor, time, citizen, percept, symptom, organ, internet, take, model, statist, carer, epoch, golden, scientif, cognit* | Includes terms from the Medical and pharmaceutical field. |
| 5 (5.1%) | *vaccin, regress, anti, myth, movement, countri, syndrom, social, incom, diagnost, affect, determin, right, iron, hcws, children, overview, diseas, court, peopl, parent, million, routin, acupunctur, danger, mortal, immun, claim, degrad, intervent* | Some terms are out of the medical field of study, and might indicate a possible discussion. |
| 6 (3.8%) | *statist, cultur, describ, infant, citat, exampl, articl, symptom, case, american, journal, metabol, combin, parent, disagr, doctor, abl, associ, construction, bordetella, basi, advers, illustr, gliadin, illusori, literatur, indic, unit, eat, rather* | A high number of terms are related to the scientometric field of study. All the terms are objective and do not indicate an opinion or a discussion. |
| 7 (4.1%) | *biomark, altmetr, nan, behavior, disord, occur, well, postpon, answer, herd, context, fear, genet, graphic, appeal, interquartil, order, vaccin, gene, sensori, evalu, modul, geneticist, chapter, lymphocyt, abstain, putat, approach, protect, homeopathi* | Includes terms from the biology, pharmacology and genetics field of study. Might also indicate a statistical analysis along with an open discussion. |
| 8 (4.9%) | *vaccin, balanc, symptom, risk, reaction, link, subgroup, mump, regress, sphere, aefi, opioid, public, prevent, record, case, food, diseas, chronic, media, claim, allerg, week, resid, advers, children, estim, strain, cobalamin, associ* | A large part of the terms are close to *WF-PUB-1998* treated thematics. Mostly from the medical field of study. |
| 9 (5.1%) | *fraud, diseas, narrat, vaccin, complic, health, controversi, comorbid, measl, polici, coliti, bowel, neurolog, travel, inflammatori, movement, trust,* | Concern the retraction phenomena, followed with some medical expressions. It also includes strong terms such as "fraud". |

| | ocean, research, retract, attribut, percept, public, futur, preserv, caus, ulcer, case, medic, problem | |
|---|---|---|
| 10 (4.9%) | vaccin, immun, misconduct, polici, patholog, retract, aefi, result, children, research, disord, qualit, report, caus, development, advanc, case, chang, internet, record, expos, mold, infecti, program, vaer, actor, live, sinc, mani, appli | General terms related to WF-PUB-1998 treated thematics, some are correlated with a discussion around the retraction phenomena. |
| 11 (5.6%) | vaccin, health, immunis, engag, communic, reason, disord, virus, diagnost, messag, coverag, examin, make, chang, accept, client, diseas, measl, development, research, consid, resist, peopl, public, evid, observ, recent, imag, effect, pervas | Terms from the medical study field related to WF-PUB-1998 treated thematics. |
| 12 (3.3%) | uncertainti, boy, scientif, gynecologist, debat, twitter, semant, ongo, intent, vaccin, disturb, variabl, messag, rhetor, liabil, frame, reddit, percept, content, sourc, gfcf, produc, paediatr, rais, pyridox, guilt, fact, advic, link, chang | Includes terms close to the computer science lexicon and from the pediatric field of study. |
| 13 (5.9%) | vaccin, incid, scientif, erad, measl, frame, viral, literatur, enceph, diseas, controversi, mother, workshop, differ, propos, expert, infect, increas, evalu, genet, dramat, recent, coalit, frequent, communic, current, link, programm, polio, scienc | Includes a large number of general terms from different fields of study, part of them are correlated with WF-PUB-1998 thematics. |

**sTable 6.** The 22 topics available in the topic model generated using the in-text citation contexts contained in the articles citing WF-PUB-1998. For each topic (row) we mention its proportion percentage in the corpus (column 1), and the 30 most relevant terms (column 2), and we give our interpretation of it (column 3).

| Topic (proportion) | Terms (the 30 most probable terms) | Interpretation |
|---|---|---|
| 1 (2.7%) | valu, reuter, thomson, worth, retract, would, impact, time, mean, paper, immun, comparison, figur, lancet, base, mening, cerebr, associ, identifi, greater, roach, regress, senior, later, europ, vitamin, viral, assign, consist, campaign | Includes few terms from the medical domain. Might talk about WF-RET-CASE, and from a statistical/mathematical perspective. Some terms include general info about the paper (metadata). |
| 2 (5.2%) | articl, control, associ, public, signific, affect, research, biopsi, scientif, follow, controversi, natur, patient, case, decad, preserv, evid, report, vaccin, use, multipl, clinic, subsequ, differ, first, cell, follicl, symptom, concern, studi | General terms which summarize what WF-PUB-1998 talks about. |
| 3 (3.9%) | articl, case, colon, development, lancet, enterocol, report, assert, associ, public, vaccin, disord, follow, signific, diagnosi, base, group, mumpsrubella, treatment, controversi, parent, scientif, evid, result, symptom, studi, publish, link, reaction, unknown | General terms which summarize what WF-PUB-1998 talks about. |
| 4 (3.9%) | articl, immun, dose, claim, lancet, requir, discredit, declin, find, infect, diseas, februari, colleagu, herd, coverag, respons, measl, report, first, sever, vaccin, regress, month, andrew, second, mump, research, countri, detail, parent | General terms which summarize what WF-PUB-1998 talks about. This topic might also outline other information related to the paper. |
| 5 (3.6%) | *vaccin, link, research, suggest, report, appear, measl, focus, scientif, develop, univers, reject, possibl, studi, associ, mump, investig, though, fund, even, continu, around, wakefieldet, newspap, result, doubt, relat, signific, high, evid* | Discusses the controversy around WF-RET-CASE. |
| 6 (2.8%) | *design, studi, expert, consider, parent, bias, receiv, requir, risk, messag, control, occur, point, factor, time, start, howev, report, third, call, know, vaccin, connect, major, qualiti, media, research, school, best, mump* | Does not include any medical term, it rather focuses on other related aspects concerned with WF-PUB-1998 |
| 7 (2.2%) | *diseas, caus, characterist, process, read, inflamm, alarm, report, assert, without, show, ileocolon, bowel, declin, safeti, student, first, exampl, intestin, media, young, detail, autist, adult, possibl, scientif, studi, even, control, wide* | General terms which summarize what WF-PUB-1998 talks about. Many terms are related to its medical background. |
| 8 (13.2%) | *regress, development, increas, hypothes, causal, vaccin, link, bowel, disord, case, report, measl, problem, symptom, author, associ, system, immun, mump, hypothesi, relationship, peptid, suggest, opioid, diseas, popul, onset, studi, risk, autist* | General terms which summarize what WF-PUB-1998 talks about. Almost all the terms are related to its medical background. |
| 9 (2.6%) | *seri, development, parent, report, abnorm, loss, child, autist, associ, acquir, consecut, articl, spectrum, coliti, pervas, symptom, specif, attent, author, count, skill, normal, public, concurr, characterist, claim, abdomin, reduc, enabl, eight* | A large number of terms from the medical domain. The connection with WF-PUB-1998 is less evident. |
| 10 (3.9%) | *mucos, regress, trigger, subtl, food, medic, symptom, development, pattern, extens, lead, clear, result, retract, disord, specif, condit, enterocol, intoler, possibl, research, also, develop, affect, suggest, case, diarrhea, includ, year, caus* | General terms which summarize *WF-PUB-1998* medical background. Might also mention its retract |
| 11 (3.7%) | *understand, paper, period, peer, cohort, development, singl, review, widespread, technic, major, help, studi, rat, scienc, public, littl, interact, relationship, origin, week, outbreak, neurochem, normal, media, articl, sometim, regress, debat, report* | Focuses on technical aspects and does not include any medical terminology. |
| 12 (10.8%) | *find, retract, uptak, studi, subsequ, media, vaccin, paper, fear, link, controversi, increas, evid, measl, scientif, articl, publish, number, safeti, mump, journal, belief, parent, public, mani, health, mother, concern, claim, lancet* | Talks and discusses WF-PUB-1998 retraction phenomena from different perspectives, e.g. social impact. It can also talk about the negative impacts. |
| 13 (3.6%) | *lancet, cost, articl, rhetor, health, paper, public, text, scienc, begin, outbreak, emerg, autist, immedi, also, featur, interpret, acquir, origin, caus, controversi, distress, depart, note, might, languag, debat, measl, behavior, vaccin* | Talks and discusses WF-RET-CASE from different perspectives, yet far from the medical domain. It might also not take in consideration the paper retraction in the discussion. |
| 14 (5.6%) | *unit, publish, studi, state, bowel, possibl, general, paper, measl, immunis, royal, case, press, diseas, report, free, group, three, kingdom, research, link, controversi, hospit, risk, would, receiv, vaccin, women, journal, earlier* | Discusses the medical conclusions arised from WF-PUB-1998 |
| 15 (3.3%) | *intestin, report, associ, studi, regress, retract, team, hospit, bowel, sever, find, ileum, royal, behavior, hypothesi, free, subsequ, caus, ethic, altmetr, problem, colleagu, vaccin, consider, research, connect, abnorm, paper, number, british* | Discusses the medical conclusions arised from WF-PUB-1998. It might point out the emerging controversies of the paper. |
| 16 (4.8%) | *associ, development, regress, whether, initi, spectrum, vaccin, bowel, trigger, possibl, propos, widespread, autoimmun, autist, disord, specif, environment, diseas, increas, coverag, concern, public, articl, hypothesi, virus, aris, question, base, andrew, sinc* | Talks about the medical thematics and conclusions of WF-PUB-1998. |
| 17 (5.4%) | *refer, articl, link, vaccin, skill, normal, histori, mump, acquir, side, bowel, support, concern, studi, public, research, suggest, associ, measl, demonstr, paper, lancet, evid, pediatr, symptom, exist, follow, prove, describ, diseas* | Discusses the conclusions and impact of WF-PUB-1998 |
| 18 (4.6%) | *receiv, countri, author, health, measl, articl, year, effect, andrew, eight, recent, parent, behavior, case, suggest, vaccin, paper, begin, safeti, british, sinc, studi, develop, short, mump, patient, campaign, publish, report, investig* | Gives an overview of WF-RET-CASE, without necessarily analysing the contents and conclusions. |
| 19 (3.2%) | *articl, factor, constip, nonspecif, britain, caus, first, autist, measl, compon, year, upon, call, intestin, publish, dquo, retract, make, immun, event, disord, neurolog, unnecessari, permeabl, find, apoptosi, vaccin, describ, scientif, syndrom* | Talks about WF-RET-CASE and the medical conclusions of the paper. |
| 20 (1.6%) | *theori, council, nationwid, queri, consequ, profession, medic, continu, ultim, despit, accept, long, case, optim, rat, vitamin, pertussi, myelogenesi, impair, persist, altern, indic, https, prsa, cross, field, coverag, notif, dismiss, collaps* | Discusses WF-PUB-1998 as a case study that might be of interest to a better understanding of the research or more specifically the medical research. |
| 21 (5%) | *vaccin, coverag, public, associ, link, articl, suggest, publish, research, media, time, citat, claim, thimeros, across, prove, particular, measl, lancet, lead, whether, parent, paper, evid, subsequ, extens, sinc, mump, around, increas* | Talks about WF-RET-CASE and consequences. Might also discuss the paper links and citations. |
| 22 (4.4%) | *connect, retract, paper, potenti, topic, disord, research, deer, receiv, exampl, studi, lancet, report, claim, elliman, causal, use, development, inflammatori, although, concern, data, type, exist, inform, bowel, publish, base, measl, attent* | Discusses WF-PUB-1998 and retraction from non-medical aspects. Might refer to it as a retraction example. |